\documentclass[aps,prl,reprint,showpacs,amsmath,amssymb]{revtex4-1}
\usepackage[usenames,dvipsnames]{xcolor}
\usepackage{graphicx}
\usepackage{natbib}
\usepackage{hyperref}
\usepackage[caption=false]{subfig} 

\newcommand{\midrule}{\hline}
\newcommand{\bottomrule}{\hline\hline}

\newcommand{\rsat}{r_{\rm sat}}

\newcommand{\NE}{N_{\rm E}}
\newcommand{\ND}{N_{\rm D}}
\newcommand{\NF}{N_{\rm F}}
\newcommand{\NP}{N_{\rm P}}

\newcommand{\vspacebeforecaption}{\vspace{-.4cm}}  
\newcommand{\vspaceaftercaption}{\vspace{-.4cm}}  

\usepackage{xcolor}

\begin{document}

\date{\today}
\title{Long-range Response in AC Electricity Grids}
\author{Daniel Jung}
\email{d.jung@jacobs-university.de}
\affiliation{
    Department of Physics \& Earth Sciences, Focus Area Health,
    Jacobs University Bremen, 28759 Bremen, Germany
}
\author{Stefan Kettemann}
\email{s.kettemann@jacobs-university.de}
\affiliation{
    Department of Physics \& Earth Sciences, Focus Area Health,
    Jacobs University Bremen, 28759 Bremen, Germany
}

\begin{abstract}

Local changes in the topology of electricity grids can cause overloads far away
from the disturbance \cite{Witthaut2013}, making the prediction of the
robustness against changes in the topology -- for example caused by power
outages or grid extensions -- a challenging task. The impact of single-line
additions on the long-range response of DC electricity grids has recently been
studied \cite{Labavic2014}. By solving the real part of the static AC load flow
equations, we conduct a similar investigation for AC grids. In a regular 2D
grid graph with cyclic boundary conditions, we find a power law decay for the
change of power flow as a function of distance to the disturbance over a wide
range of distances. The power exponent increases and saturates for large system
sizes. By applying the same analysis to the German transmission grid topology,
we show that also in real-world topologies a long-ranged response can be found.

\end{abstract}



\maketitle



Power grids reliably provide eletrical power to billions of individuals. For
example, in Germany the average outage time experienced by a consumer in 2006
was 20 minutes and continued to decrease in the last decade to 12.5 minutes in
2014 \cite{bundesnetzagentur}. Still, the energy transition towards an
increased supply of decentralized renewable energy raises concerns that the
change from the previously centralised power production with unidirectional
power flow towards a decentralised electrical power system with bidirectional
flow might be harmful for the stability of electricity grids.
In a conventional energy grid, the largest consumers (industry) as well as the
largest generators usually consist of large rotating masses, so small
perturbations are sufficiently damped. As of the energy transition towards an
increasingly decentral power generation, generators that do not possess this
kind of buffer of electrical energy in form of inertia -- namely photovoltaic
cells -- produce an increasing share of the energy supply.
Disturbances and power outages might spread more easily in highly connected
grids and cause nonlocal disturbances, which may cause larger instabilities of
the entire grid.
Therefore, it is essential to get a better understanding of the physical
mechanisms leading to nonlocal disturbances and how their spreading depends on
the connectivity and topology of the grid.

In a previous work, the long-range response to the addition of a single
transmission line has been studied for the case of a DC grid including losses
by Joule heating \cite{Labavic2014}. It has been demonstrated that the absolute
change of transmitted power as function of the distance to the perturbation
follows a power law. Real power transmission grids usually use three-phasic
alternating current (AC) \cite{Heuck2013}. Hence, we are going to perform a
similar analysis for AC networks.


\begin{figure}
    \begin{center}
        \includegraphics[width=.6\columnwidth]{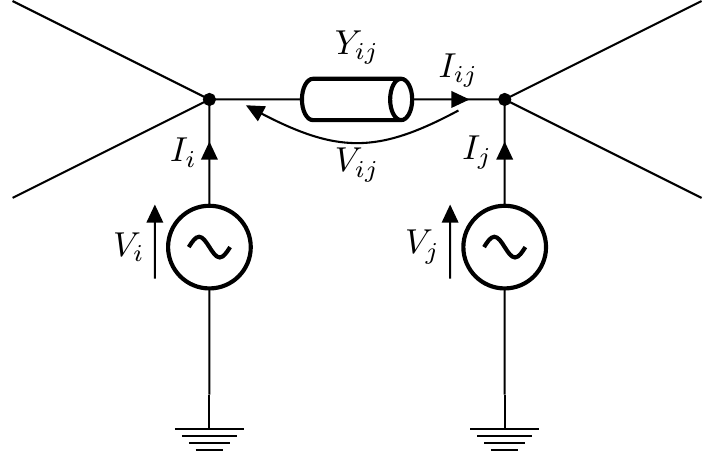}
    \end{center}
    \caption{
        Sketch of a section of an electricity grid with a transmission line
        connecting nodes $(i,j)$ with admittance $Y_{ij}$, carrying a current
        $I_{ij}$ at voltages $V_i$ and $V_j$.
     }
    \label{fig:ac-grid}
\end{figure}

\begin{figure}
    \begin{center}
        \subfloat[\label{fig:flow-phase}]{
            \includegraphics[width=.5\columnwidth]{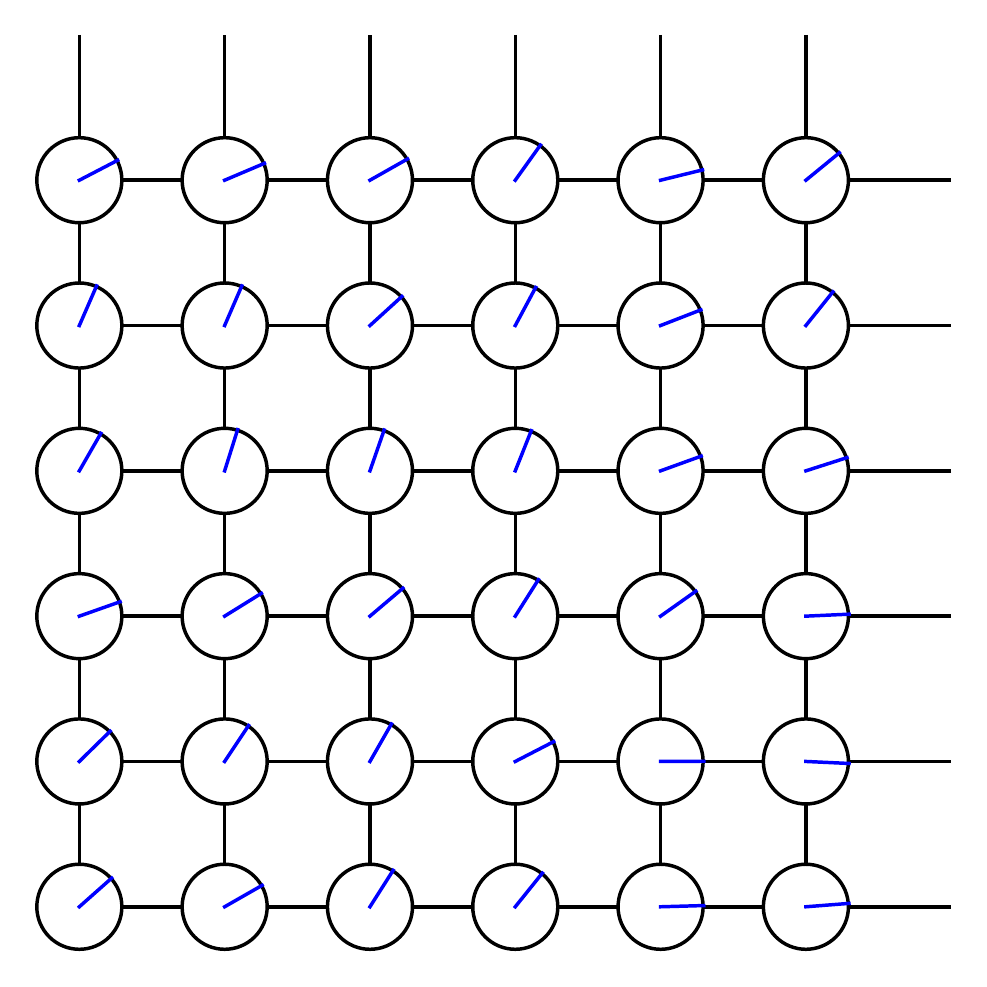}
        }
        \subfloat[\label{fig:flow-flow}]{
            \includegraphics[width=.5\columnwidth]{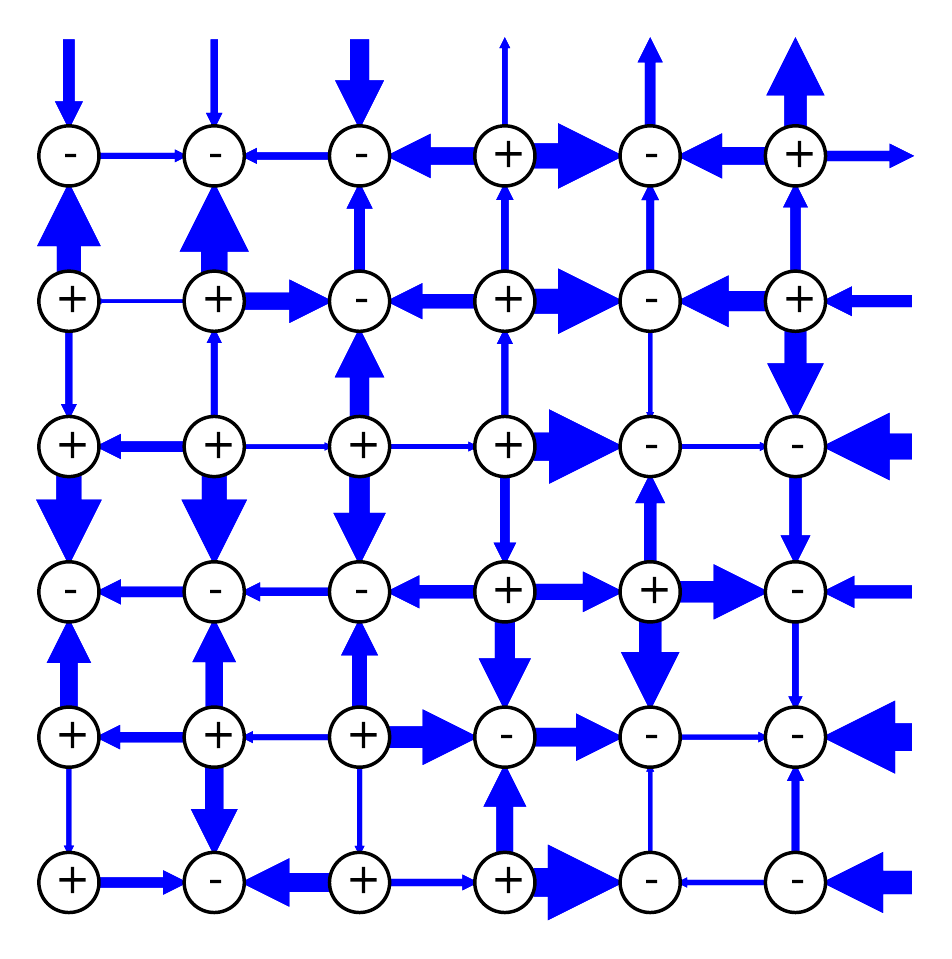}
        }
    \end{center}
    \vspacebeforecaption
    \caption{
        \protect\subref{fig:flow-phase} Voltage phase distribution of an
        example system with binary distribution of the net generated power $P_i
        \in \{-P,P\}$ at each node.
        \protect\subref{fig:flow-flow} The respective power flow transmitted
        through each transmission line. The size of the arrows is proportional
        to the transmitted power $F_{ij}$. The symbols ``+'' and ``-'' indicate
        generators and consumers, respectively.
    }
    \vspaceaftercaption
    \label{fig:flow}
\end{figure}

An electrical power transmission system can formally be described as a
multigraph $G$ that consists of nodes $i,j \in \mathcal{N}$ (substations) and
edges $(i,j) \in \mathcal{E}$ (transmission lines), where $\mathcal{N}$ is the
set of all nodes and $\mathcal{E}$ the set of all edges of $G$. The network
model gains its physical meaning by defining appropriate node and edge
attributes. Starting from the \emph{Kirchhoff's laws} and \emph{Ohm's law}, it
is straighforward to write down the \emph{steady state power flow equations}
for a three-phase AC network,
\begin{equation}
    S_i - 3 V_i \sum_j Y_{ij}^* (V_i - V_j)^* = 0 \quad\rm,
    \label{eq:powflow}
\end{equation}
where the three alternating currents of the same frequency are phase-shifted by
120 degrees. Here, $S_i = P_i + i Q_i$ is the \emph{net generated power}
entering the network at node $i$ (negative for a consumer) and $V_i$ is the
terminal voltage of the ``machine'' connected at node $i$. $Y_{ij}$ is the
admittance of the transmission line $(i,j)$ \cite{conj}. For an arbitrary
transmission line in an (one-phasic) AC network, the labeling is illustrated in
Fig.~\ref{fig:ac-grid}.

If we restrict ourselves to a purely inductive grid,  
\begin{equation}
    Y_{ij} = \frac{1}{i \omega L_{ij}} \quad\rm,
    \label{eq:admittance}
\end{equation}
we can assume sinusoidal voltages with a constant magnitude $|V_i| \equiv V$
\cite{Kettemann2015},
\begin{equation}
    V_i(\omega, t) = V e^{i \varphi_i(\omega, t)} \quad\rm,
    \label{eq:voltage}
\end{equation}
with phase angles $\varphi_i(\omega, t) = \omega t + \theta_i(t)$. Thus, only
the phase difference between adjacent nodes gives rise to an electrical
current. We are looking for steady states with constant grid frequency $\omega$
(e.g. $\omega = 2\pi \cdot 50\,\text{Hz}$). The \emph{power capacity} of a
transmission line (a set of three wires) is given by
\begin{equation}
    K_{ij} = \frac{3 V^2}{\omega L_{ij}} \quad\rm.
    \label{eq:powcap}
\end{equation}
So the power flow equations \eqref{eq:powflow} become
\begin{equation}
    S_i =
    i \sum_j K_{ij}\, \left(1 - e^{i (\theta_i - \theta_j)} \right) \quad\rm.
    \label{eq:powflow2}
\end{equation}
The real part gives the balance of the \emph{active power} flow,
\begin{equation}
    P_i = \sum_j K_{ij} \sin(\theta_i - \theta_j) \quad\rm,
    \label{eq:powflowreal}
\end{equation}
which determines the phases $\theta_i$ at all nodes for the given load
distribution $P_i$. The imaginary part of Eq.~\eqref{eq:powflow2} gives the
balance of the \emph{reactive power} $Q_i$. For the constant voltage amplitudes
considered in this article, the $N$ active power equations determine the $N$
phases $\varphi_i$ which also fix the reactive power $Q_i$ at all nodes $i$.
Eq.~\ref{eq:powflowreal} can also be derived from the established
\emph{synchronous motor model} (swing equation)
\cite{Filatrella2008,Nishikawa2015} by setting all time-dependent terms to
zero.

If there exists a stationary solution $\theta_i$ for the chosen system
parameters, it can easily be found using a standard root-finding algorithm
\cite{root,multistability} (see Fig.~\ref{fig:flow-phase} for an example phase
distribution).  Given a solution for the phase distribution $\theta_i$, the
transmitted power $F_{ij}$ from node $i$ to node $j$ is given by
\begin{equation}
    F_{ij} = K_{ij} \sin(\theta_i - \theta_j) \quad\rm,
    \label{eq:powtransmit}
\end{equation}
as plotted in Fig.~\ref{fig:flow-flow}, where the thickness of the arrows is
proportional to $|F_{ij}|$.

\begin{figure}
    \begin{center}
        \subfloat[\label{fig:flow2-flow}]{
            \includegraphics[width=.5\columnwidth]{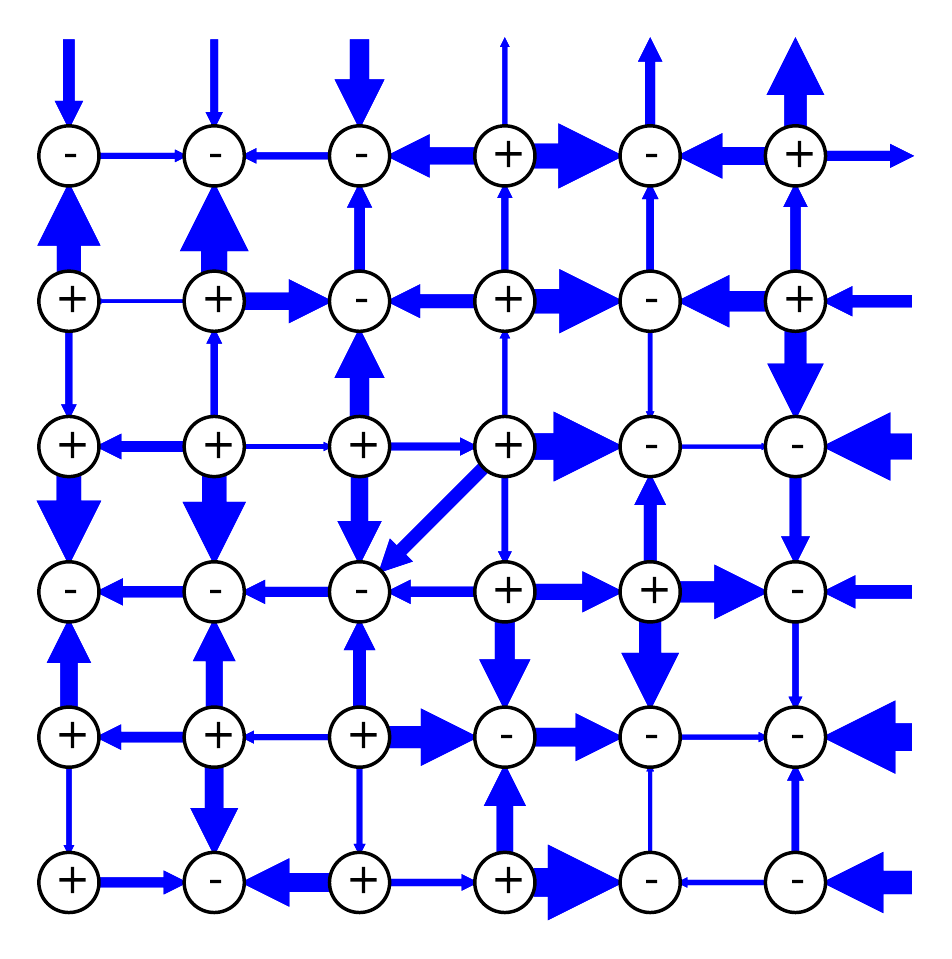}
        }
        \subfloat[\label{fig:flow2-diff}]{
            \includegraphics[width=.5\columnwidth]{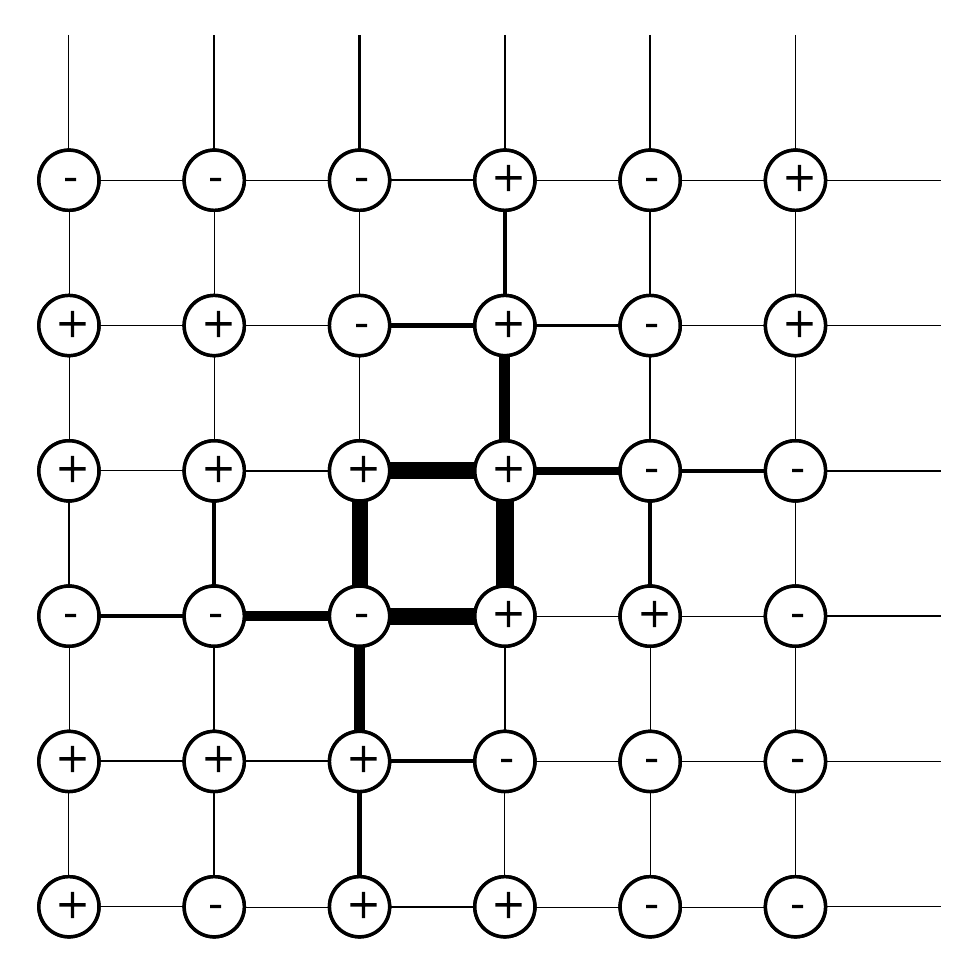}
        }
    \end{center}
    \vspacebeforecaption
    \caption{
        \protect\subref{fig:flow2-flow} Power flow of the same system as in
        Fig.~\ref{fig:flow} after the addition of another transmission line.
        \protect\subref{fig:flow2-diff} Absolute change of the power flows
        $|\Delta F_{ij}|$ in the original grid after adding the line. The width
        of the lines is proportional to $|\Delta F_{ij}|$.
    }
    \vspaceaftercaption
    \label{fig:flow2}
\end{figure}

Then we add a single line to the graph and calculate the modified transmitted
powers $F'_{ij}$ (see Fig.~\ref{fig:flow2-flow}). We calculate the difference
$\Delta F_{ij} = F'_{ij} - F_{ij}$ and study the average absolute difference
$\langle | \Delta F_{ij} | \rangle$ of all edges as a function of their
distance to the added line (see Fig.~\ref{fig:flow2-diff}). The average covers
all edges with the same distance $r$ to the added line as well as an ensemble
average over $R = 1000$ random load distributions $P_i$.

\begin{figure}
    \begin{center}
        \includegraphics[width=.7\columnwidth]{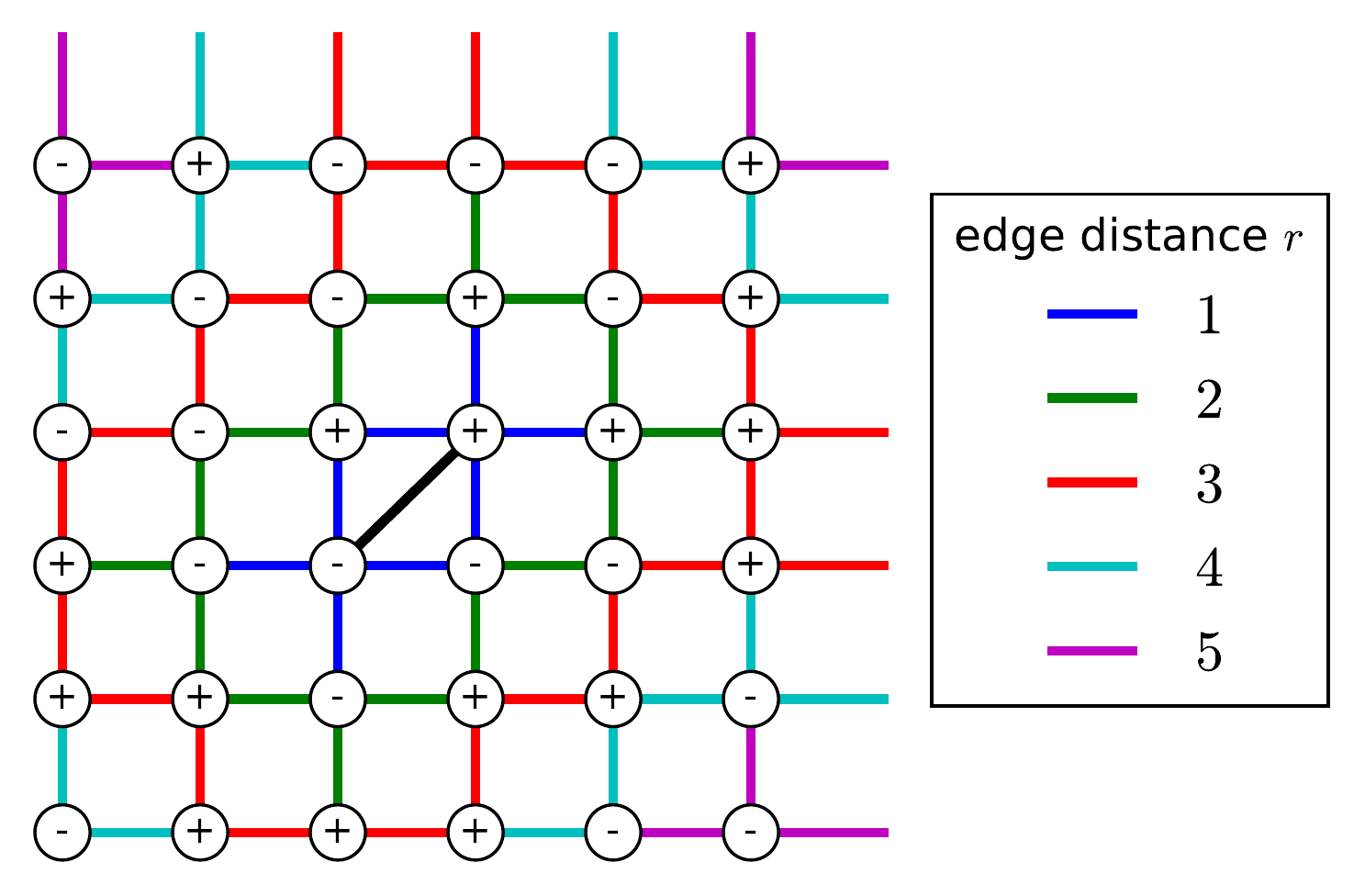}
    \end{center}
    \vspacebeforecaption
    \caption{
        (Color online) Classification of the transmission lines by their
        distance $r$ to the added line (black).
    }
    \vspaceaftercaption
    \label{fig:distance}
\end{figure}

We first consider a regular 2D grid graph of linear size $L$ with cyclic
boundary conditions, so the system size (number of nodes) is given by $N =
L^2$.  In this graph it is particularly simple to define a measure for the
distance between two edges, by counting the number of edges that have to be
passed (shortest path), as illustrated in Fig.~\ref{fig:distance}. We further
consider a binary distribution for the nodal \emph{net generated power} $P_i$,
where nodes with $P_i = +P$ are regarded as generators and nodes with $P_i =
-P$ as consumers. We set the power capacity of all lines to $K_{ij} = K$, so we
can note down all power quantities in units of $K$. In order to find a stable
solution, the condition $\sum_i P_i = 0$ must be fulfilled at all times, which
rules out the possibility of odd linear system sizes $L$.

In order to precisely control the amount of randomness in the system, we use
the following procedure to generate a random distribution of the $P_i$:  We
start from a periodic arrangement of generators and consumers \cite{antiferro}
and divide the graph into two subgraphs, one carrying all $N/2$ generators and
the other all $N/2$ consumers. Then, $p$ different nodes are chosen randomly
from each subgraph, forming $p$ generator-consumer pairs. Finally, each of
these generator-consumer pairs is swapped. By generating a permutation of the
periodic arrangement in this way, it is ensured that no node is swapped twice,
and the degree of randomness $w \in [0,1]$ can precisely be specified as
\begin{equation}
    w = \frac{4p}{N} \quad\rm,
    \label{eq:randomness}
\end{equation}
depending on the number of swapped generator-consumer pairs $p$. The reasoning
here is that after $p=N/2$ permutations, the periodic arrangement is reached
again, but with all original consumers and generators swapped. If only half as
many permutations are done, $p_{\rm max}=N/4$, the state should be the furthest
away from one of the two possible periodic arrangements and thus correspond to
maximum randomness. There is a finite number of possible realizations $P_i$,
given by the ensemble size
\begin{equation}
    \NE = \binom{N/2}{p} \quad\rm,
    \label{eq:ensemblesize}
\end{equation}
which is large compared to $R$, even for the smallest considered system sizes.

To study the long-range response of the grid to the added transmission line, we
classify all edges of the graph by their distance $r$ to the added edge by
counting the number of edges that link the considered edge to the added edge
(see Fig.~\ref{fig:distance} for an illustration). We average the change in the
amplitude of the transmitted power $|\Delta F_{ij}|$ over all edges with the
same distance $r$ to the added line, and perform an ensemble average $\langle
|\Delta F_{ij}| \rangle (r)$ over $R = 1000$ realizations with the same
randomness $w$. Realizations for which no steady state solution can be found
are skipped. This happens in particular if the transmitted power $F_{ij}$ is
exceeding the capacity $K_{ij}$ of at least one transmission line $(i,j)$, so
that Eq.~\eqref{eq:powtransmit} cannot be fulfilled.
For the square grid, the power capacity $K$ has to exceed the critical value
$K_{\rm c} = P / 4$ in order to obtain a steady state power flow at all
\cite{kc}. Since for a random arrangement of $P_i$, clusters of $N_{\rm c}$
generators or $N_{\rm c}$ consumers can occur with total power $N_{\rm c} P$,
the lines connected to these clusters can be overloaded even if the power
capacity exceeds the critical value $K > K_{\rm c}$, and no steady state
solution is found. Thus, only a certain subset of realizations leads to a
solution for the phase distribution $\theta_i$.

\begin{figure}
    \begin{center}
        \includegraphics[width=\columnwidth]{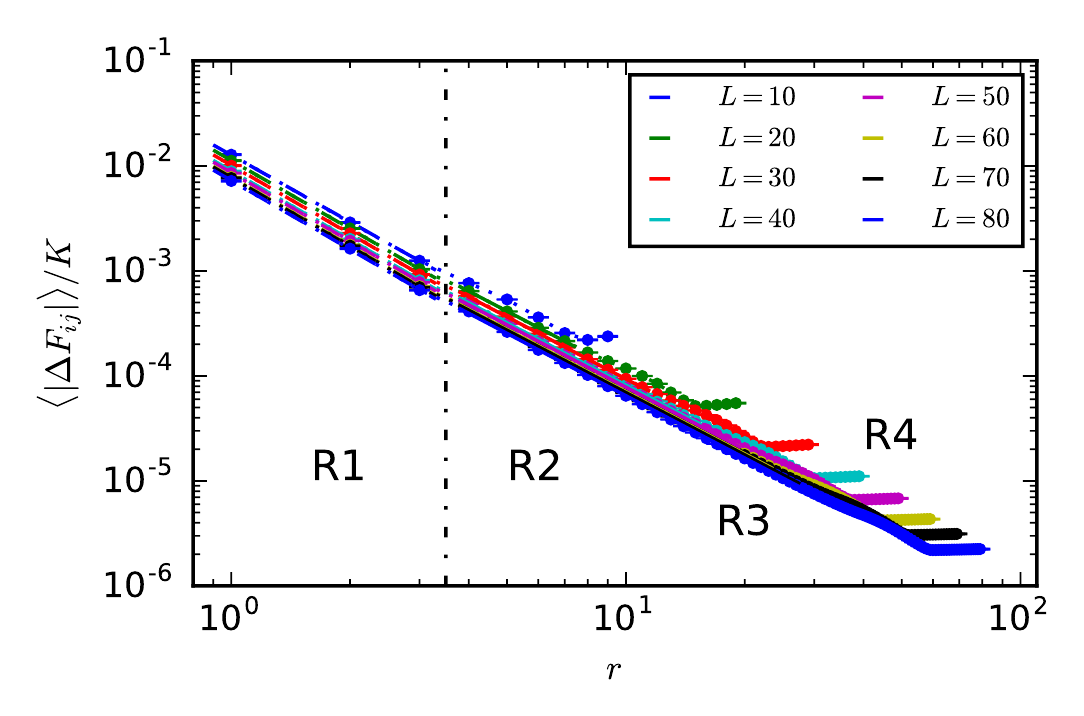}
    \end{center}
    \vspacebeforecaption
    \caption{
        (Color online) Double-logarithmic plot of $\langle | \Delta F_{ij} |
        \rangle (r)$ for different system sizes $L$ and $P/K = 0.25$. For the
        regimes R1 and R2, the data has been fitted to a power law. The error
        bars correspond to a $95\,\text{\%}$ confidence level
        \cite{confidence}.
    }
    \vspaceaftercaption
    \label{fig:scaling}
\end{figure}

\begin{figure}
    \includegraphics[width=.8\columnwidth]{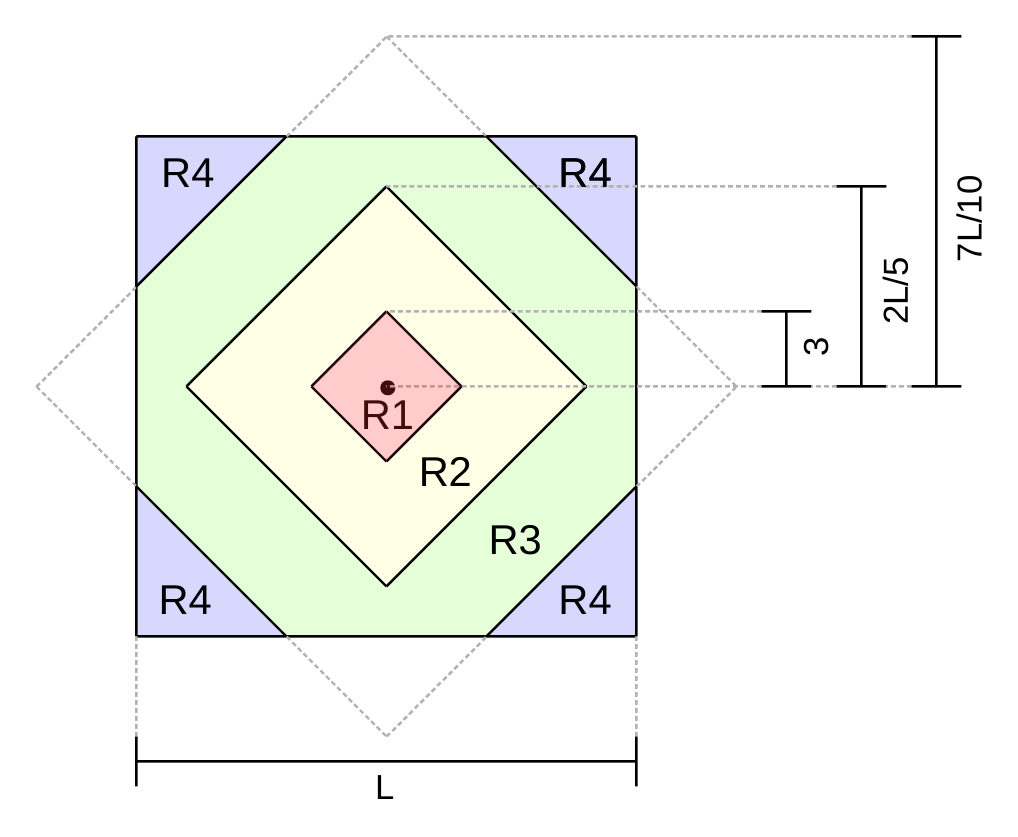}
    \vspacebeforecaption
    \caption{
        (Color online) Sketch of the distance regions R1, R2, R3 and R4 in a
        cyclic $L \times L$ 2D grid for which different long-range behaviors
        are observed. The location of the topological perturbation (added
        transmission line) is marked by the black dot in the center.
    }
    \vspaceaftercaption
    \label{fig:grid2d-regions}
\end{figure}

\begin{figure}
    \begin{center}
        \includegraphics[width=\columnwidth]{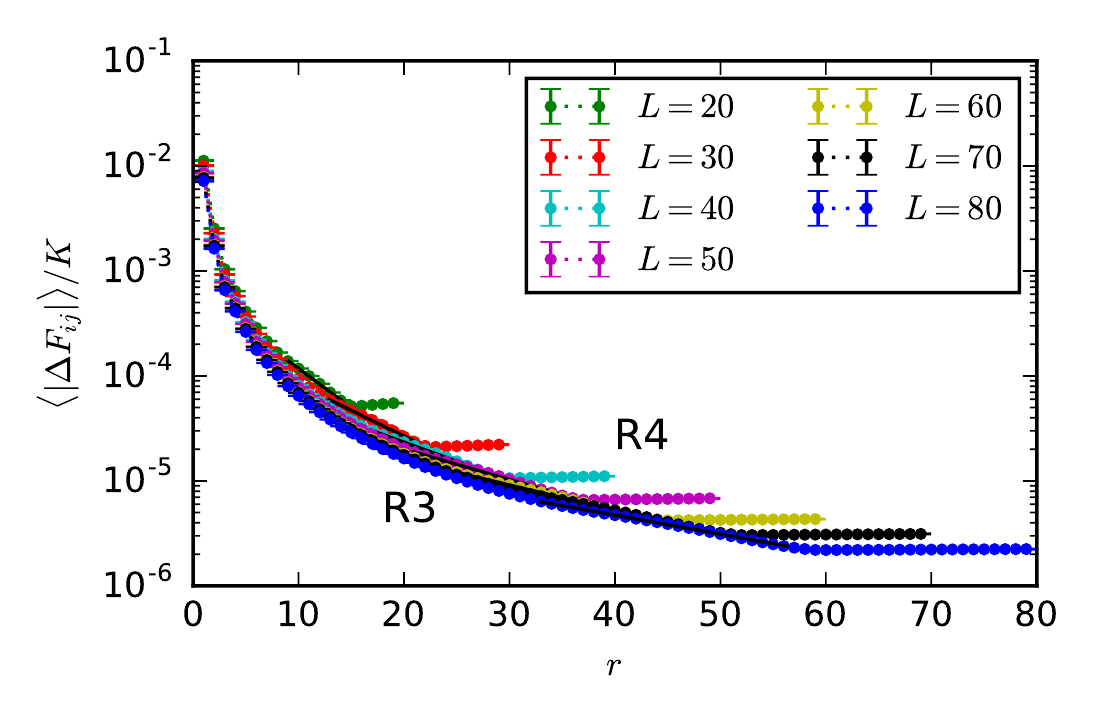}
    \end{center}
    \vspacebeforecaption
    \caption{
        (Color online) Semi-logarithmic plot of $\langle | \Delta F_{ij} |
        \rangle (r)$ for different system sizes $L$ and $P/K = 0.25$.  The data
        belonging to the regime R3 has been fitted to an exponential decay law
        \eqref{eq:explaw}. The error bars correspond to a $95\,\text{\%}$
        confidence level \cite{confidence}.
    }
    \vspaceaftercaption
    \label{fig:scaling-r3}
\end{figure}

$\langle | \Delta F_{ij} | \rangle (r)$ has been calculated for different
linear system sizes $L$ and plotted on a double-logarithmic scale in
Fig.~\ref{fig:scaling}. Up to a certain distance $r_{\rm sat}$ (saturation
distance), $\langle | \Delta F_{ij} | \rangle$ is a steadily decreasing
function of $r$. The value $r_{\rm sat}$ is found to depend linearly on the
linear system size $L$, roughly as $r_{\rm sat} = 7L/10$. For distances $r$
exceeding $r_{\rm sat}$ (region R4), the data saturates, or even a small
increase is noticable, which is caused by the cyclic boundary conditions as
confirmed in a comparative analysis. For distances $r < r_{\rm sat}$ we are
able to identify two regimes R1 ($1 \leq r \leq 3$) and R2 ($4 \leq r \leq
2L/5$) that both show a power-law-like behavior of $\langle | \Delta F_{ij} |
\rangle (r)$, but with different power exponents $b$. Beyond R2 (for about $r >
2L/5$), there is a region where the data is closer to an exponential law
(region R3), as the semi-logarithmical plot in Fig.~\ref{fig:scaling-r3} shows.
The four regions of the 2D grid with different dependencies of $\langle |
\Delta F_{ij} | \rangle$ on $r$ are sketched in Fig.~\ref{fig:grid2d-regions}.

\begin{figure}
    \begin{center}
        \includegraphics[width=\columnwidth]{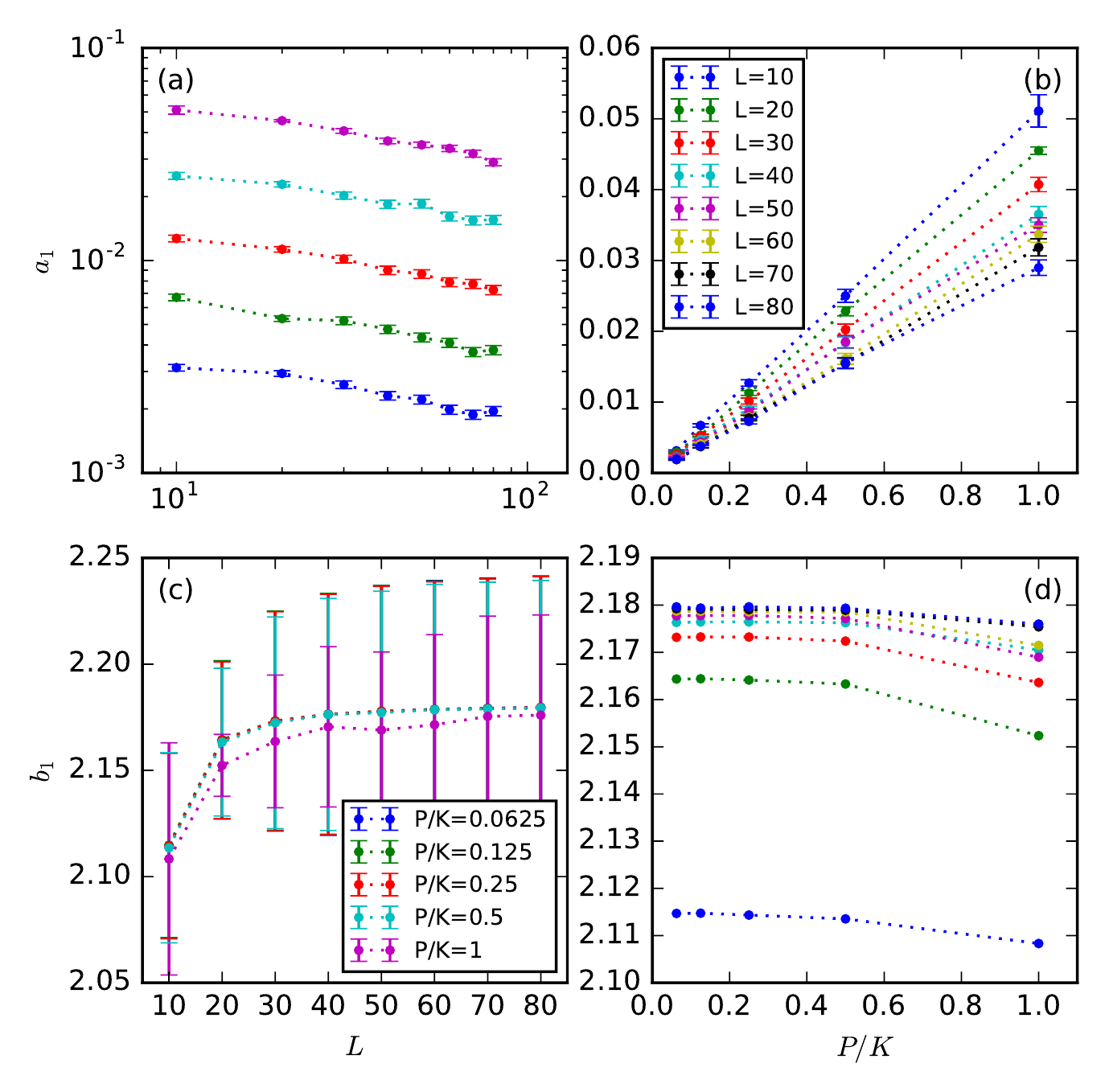}
    \end{center}
    \vspacebeforecaption
    \caption{
        Dependence of the fit parameters (a/b) $a_1$ and (c/d) $b_1$ (regime
        R1) on (a/c) the linear system size $L$ and (b/d) the ratio $P/K$. For
        better readability, the error bars are not shown in (d). Otherwise, the
        error bars correspond to a $95\,\text{\%}$ confidence level
        \cite{confidence}.
    }
    \vspaceaftercaption
    \label{fig:fp1}
\end{figure}

\begin{figure}
    \begin{center}
        \includegraphics[width=\columnwidth]{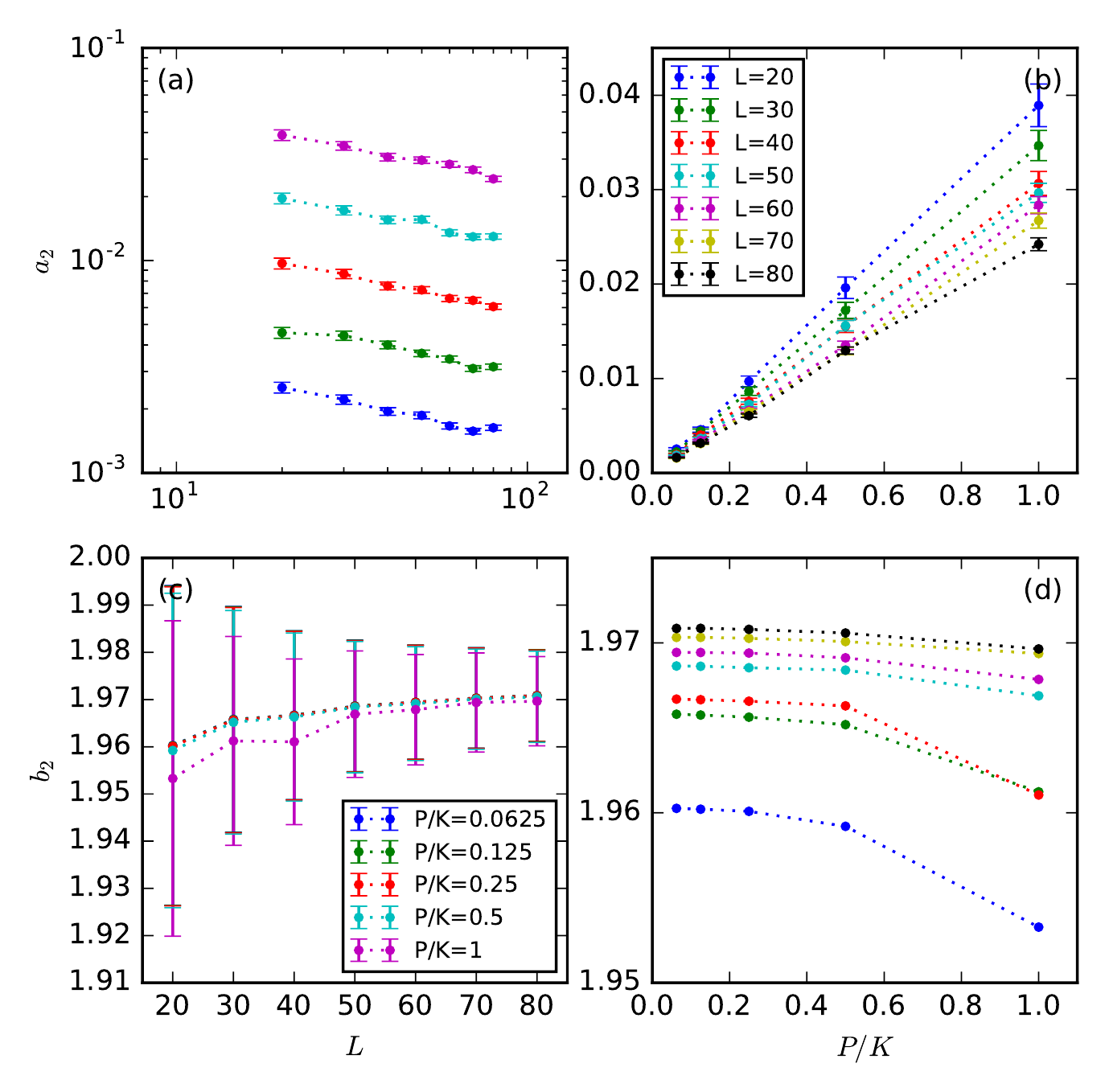}
    \end{center}
    \vspacebeforecaption
    \caption{
        Dependence of the fit parameters (a/b) $a_2$ and (c/d) $b_2$ (regime
        R2) on (a/c) the linear system size $L$ and (b/d) the ratio $P/K$. For
        better readability, the error bars are not shown in (d). Otherwise, the
        error bars correspond to a $95\,\text{\%}$ confidence level
        \cite{confidence}.
    }
    \vspaceaftercaption
    \label{fig:fp2}
\end{figure}

For the regimes R1 and R2, we fit the data to a power law, using the fit model
\begin{equation}
    f(r) = a_k r^{-b_k} \quad\rm,
    \label{eq:powerlaw}
\end{equation}
where $k$ is either 1 (regime R1) or 2 (regime R2).  The fit results for regime
R2 and $P/K=0.25$ are listed in Tab.~\ref{tab:scaling2a} (see the appendix for
more complete list of fit results). To assess the quality of the fit, we
compute the $\chi^2$ statistic and the goodness of fit probability $Q$
\cite{gof}. For the error bars, a $95\,\%$ confidence level is considered
\cite{confidence}.

\begin{table}[bh]
    \centering
    \caption{
        Fit results for regime R2 and $P/K=0.25$ (see Tab.~\ref{tab:scaling2}
        in the appendix for other $P/K$). For each parameter $L$, only data for
        $4 \leq r \leq (2L/5)-1$ is considered. To assess the quality of the
        fit, $\chi^2$ and the quality of fit probability $Q$ are given, along
        with the number of data points $\ND$ and the number of degrees of
        freedom $\NF$ \cite{gof}.
    }
    \begin{tabular}{cccrrrl}
    \toprule
     $L$ &          $a_2$ &            $b_2$ &  $N_{\rm D}$ &  $N_{\rm F}$ &  $\chi^2$ &                   $Q$ \\
    \midrule
      20 &  $9.69\pm0.29$ &  $1.960\pm0.017$ &            4 &            2 &    $7.68$ &   $2.2\cdot 10^{ -2}$ \\
      30 &  $8.64\pm0.22$ &  $1.966\pm0.012$ &            8 &            6 &   $63.89$ &   $7.3\cdot 10^{-12}$ \\
      40 &  $7.59\pm0.16$ &  $1.967\pm0.009$ &           12 &           10 &  $150.37$ &   $3.1\cdot 10^{-27}$ \\
      50 &  $7.27\pm0.13$ &  $1.969\pm0.007$ &           16 &           14 &  $250.56$ &   $2.2\cdot 10^{-45}$ \\
      60 &  $6.63\pm0.11$ &  $1.969\pm0.006$ &           20 &           18 &  $389.03$ &   $1.8\cdot 10^{-71}$ \\
      70 &  $6.49\pm0.10$ &  $1.970\pm0.005$ &           24 &           22 &  $549.15$ &  $3.9\cdot 10^{-102}$ \\
      80 &  $6.06\pm0.09$ &  $1.971\pm0.005$ &           28 &           26 &  $644.71$ &  $2.7\cdot 10^{-119}$ \\
    \bottomrule
    \end{tabular}
    \label{tab:scaling2a}
\end{table}

Due to the small number of data points ($\ND=3$) in the small distance regime
R1, the fit quality in regime R1 cannot be expected to be acceptable. But also
in regime R2, where -- depending on the system size -- a much larger number of
data points is accessible, the values for the \emph{goodness of fit
probability} $Q$ \cite{gof} are very low, indicating that the data cannot be
described by a pure power law, despite the obvious qualitative indications.
On the one hand, this is due to the crossover from one region to another.
On the other hand, in many cases the data seems to contain small oscillations
around the fitted power law, a behavior our fit model is not able to cover.
Nevertheless, we find clear evidence that the response of the power flow in an
AC power transmission grid to a local modification of the topology is of
long-ranged nature, decaying mainly with a power law with distance $r$.

In Figs.~\ref{fig:fp1} and \ref{fig:fp2}, we analyze the dependence of the fit
parameters $a_1$, $b_1$ (regime R1) and $a_2$, $b_2$ (regime R2) on the linear
system size $L$ and the power transmission capacity ratio $P/K$.  The data
suggests a saturation of the power exponents $b_1$ and $b_2$ in the limit of
large system sizes. We find the largest $b_1 = 2.180 \pm 0.031$ in R1 and $b_2
= 1.971 \pm 0.005$ in R2, for a linear system size of $L = 80$, when the power
is set to $P/K = 0.25$. We also detect a clear dependence on the ratio $P/K$:
$b_1$ and $b_2$ both show a decrease when $P/K$ is increased, leading to an
even more pronounced long-range behavior.  This decrease is expected, since the
closer the electricity grid is driven to its maximal transmission capacity
(some critical ratio $(P/K)_{\rm c}$ that depends on the topology and load
distribution), the smaller are the allowed changes $\Delta F_{ij}$ when adding
the transmission line.

In contrast to the power exponents $b_k$, the preexponentials $a_k$ in
Eq.~\eqref{eq:powerlaw} increase linearly with $P/K$, as shown in
Figs.~\ref{fig:fp1}b and \ref{fig:fp2}b. This behavior is expected as well, as
the addition of a transmission line with power capacity $K$ will lead to a
redistribution of the power flow in proportion to the generated power $P/K$, so
that $\langle |\Delta F_{ij}| \rangle \sim P/K$ is expected on average.
Further, $a_1$ and $a_2$ decrease monotonically, roughly following a power law,
with linear system size $L$ (see double logarithmic plots in
Figs.~\ref{fig:fp1}a and \ref{fig:fp2}a).

For larger distances, beyond the region R2, no power law behavior can be
observed anymore.  Ignoring also here a certain inaccuracy due to crossover
effects and oscillating data, one can however roughly identify an exponential
decay in region R3, as illustrated by the semi-logarithmic representation of
the data in Fig.~\ref{fig:scaling-r3}.  The data in regime R3, stretching from
$r=2L/5+1$ to $r=7L/10$, is fitted to an exponential decay law,
\begin{equation}
    f_3(r) = g_3 \exp(-h_3 r) \quad\rm.
    \label{eq:explaw}
\end{equation}
The resulting fit parameters are listed in the appendix in
Tab.~\ref{tab:scaling3}.

\begin{figure}
    \begin{center}
        \includegraphics[width=\columnwidth]{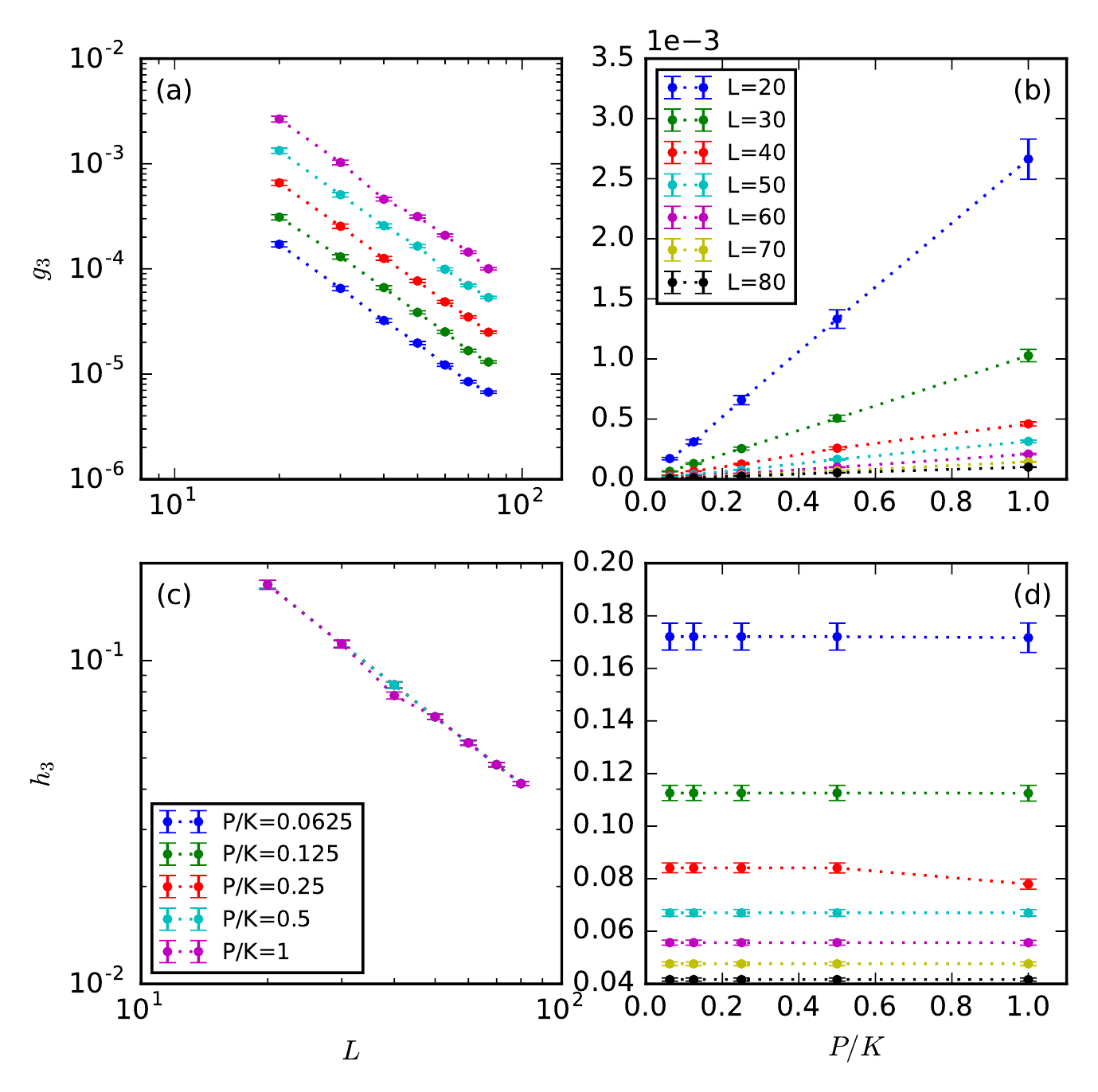}
    \end{center}
    \vspacebeforecaption
    \caption{
        Dependence of the fit parameters (a/b) $g_3$ and (c/d) $h_3$ (regime
        R3) on (a/c) the linear system size $L$ and (b/d) the ratio $P/K$. The
        error bars correspond to a $95\,\text{\%}$ confidence level
        \cite{confidence}.
    }
    \vspaceaftercaption
    \label{fig:fp3}
\end{figure}

The dependence of the fit parameters $g_3$ and $h_3$ on the linear system size
$L$ and the power ratio $P/K$ are visualized in Fig.~\ref{fig:fp3}.  The decay
of the preexponential $g_3$ with increasing linear system size $L$ is even
better described by a power law as that of the fit parameters $a_1$ and $a_2$
in regions R1 and R2 (see Fig.~\ref{fig:fp3}a), and increases linearly with
increasing $P/K$, just as $a_1$ and $a_2$ (see Fig.~\ref{fig:fp3}b).  The
exponent $h_3$ decreases as a power law with increasing $L$ as well, but stays
constant for increasing $P/K$ (see Figs.~\ref{fig:fp3}c+d).

\begin{figure}
    \includegraphics[width=\columnwidth]{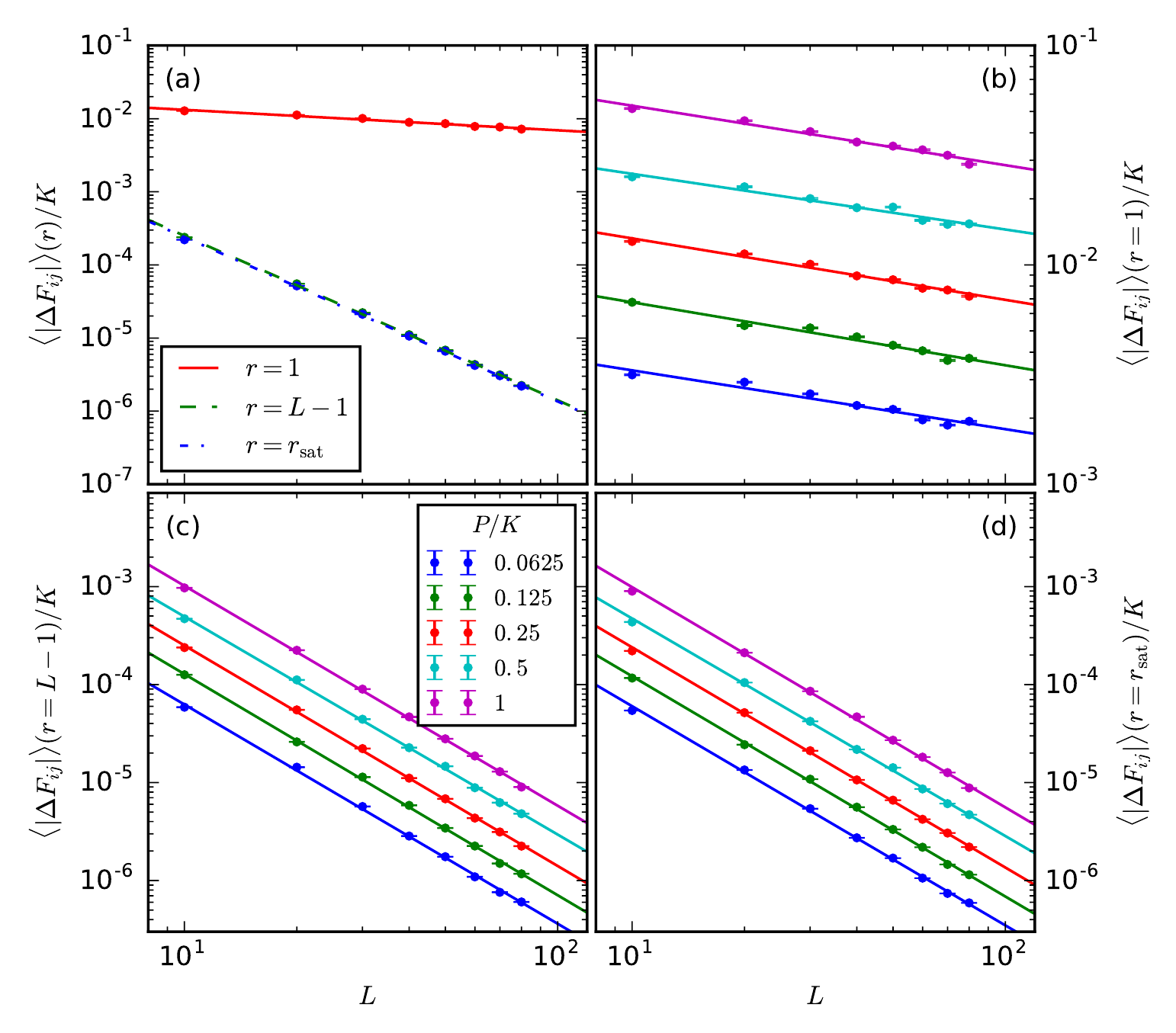}
    \vspacebeforecaption
    \caption{
        (Color online) $\langle |\Delta F_{ij}| \rangle$(r) for (b) $r=1$
        (smallest distance), (c) $r=L-1$ (largest distance) and (d) $r=\rsat$
        (saturation distance) as a function of linear system size $L$. (a) All
        of the above for $P/K=0.25$. The error bars correspond to a
        $95\,\text{\%}$ confidence level \cite{confidence}.
    }
    \vspaceaftercaption
    \label{fig:fss-F}
\end{figure}

We also analyze the value of $\langle |\Delta F_{ij}| \rangle (r)$ directly as
function of linear system size $L$ for the smallest distance $r=1$, the largest
distance $r=L-1$ and at the value $r = \rsat$, where $\langle |\Delta F_{ij}|
\rangle$ starts to saturate, as shown in Fig.~\ref{fig:fss-F}. The dependence
on $L$ is qualitatively following a power law, $g(L) = e + c L^{-d}$, in
accordance with the behavior of the parameters $a_1$, $a_2$ and $g_3$. The fit
results are summarized in Tab.~\ref{tab:fss-F} and visualized in
Fig.~\ref{fig:fss-F}, setting $e = 0$. We find a different power exponent $d$
in the small distance regime at $r=1$ than in the limit of large distances
$r=L-1$. The low fit quality again shows that the data does not follow a pure
power law (with $e = 0$) \cite{underestimated}.

Having verified the long-range response of local grid modifications in AC
electricity grids, a comparison of the results to an earlier study considering
long-range response in DC grids \cite{Labavic2014} is in place, where a power
capacity ratio of $P/K=0.1$ has been used. In the DC case, also two power-law
regimes R1 and R2 have been found, and also there, the power exponent $b_1^{\rm
DC}$ is larger than $b_2^{\rm DC}$.  The absolute values of the exponents $b_k$
are however different in the two studies: In the short distance regime (R1),
for a linear system size of $L = 50$, a value of $b_1^{\rm DC} = 2.08$ is found
\cite{Labavic2014}, while for the AC model and $P/K=0.125$, we find $b_1 =
2.178 \pm 0.030$. In the medium distance regime (R2), the DC study finds
$b_2^{\rm DC} = 1.32$ \cite{Labavic2014}, while we find a remarkably different
value of $b_2 = 1.969 \pm 0.007$.

\begin{table}
    \centering
    \caption{
        (a) Fit results for the finite size scaling of $\langle |\Delta F_{ij}|
        \rangle$(r) for $r=1$ (smallest distance), $r=L-1$ (largest distance)
        and $r=\rsat$ (saturation distance) as a function of linear system size
        $L$. To assess the quality of the fit, $\chi^2$ and the quality of fit
        probability $Q$ are given \cite{gof}, along with the number of data
        points $\ND$.
    }
    \begin{tabular}{llrllll}
    \toprule
               $r$ &     $P/K$ &  $N_{\rm D}$ &              $c$ &            $d$ &  $\chi^2$ &          $Q$ \\
    \midrule
               $1$ &  $0.0625$ &            8 &  $0.006\pm0.001$ &  $0.27\pm0.02$ &   $254.2$ &   $10^{-51}$ \\
               $1$ &  $0.1250$ &            8 &  $0.013\pm0.001$ &  $0.29\pm0.02$ &   $151.6$ &   $10^{-29}$ \\
               $1$ &  $0.2500$ &            8 &  $0.025\pm0.001$ &  $0.28\pm0.01$ &    $87.4$ &   $10^{-16}$ \\
               $1$ &  $0.5000$ &            8 &  $0.047\pm0.003$ &  $0.25\pm0.02$ &   $158.5$ &   $10^{-31}$ \\
               $1$ &  $1.0000$ &            8 &  $0.099\pm0.006$ &  $0.27\pm0.02$ &   $113.2$ &   $10^{-21}$ \\
             $L-1$ &  $0.0625$ &            8 &  $0.011\pm0.001$ &  $2.24\pm0.03$ &   $430.1$ &   $10^{-89}$ \\
             $L-1$ &  $0.1250$ &            8 &  $0.023\pm0.002$ &  $2.26\pm0.02$ &   $213.0$ &   $10^{-42}$ \\
             $L-1$ &  $0.2500$ &            8 &  $0.044\pm0.003$ &  $2.25\pm0.02$ &   $194.1$ &   $10^{-38}$ \\
             $L-1$ &  $0.5000$ &            8 &  $0.082\pm0.007$ &  $2.22\pm0.02$ &   $274.9$ &   $10^{-56}$ \\
             $L-1$ &  $1.0000$ &            8 &  $0.178\pm0.014$ &  $2.24\pm0.02$ &   $218.6$ &   $10^{-44}$ \\
     $r_{\rm sat}$ &  $0.0625$ &            8 &  $0.010\pm0.001$ &  $2.24\pm0.03$ &  $1533.6$ &       $0.00$ \\
     $r_{\rm sat}$ &  $0.1250$ &            8 &  $0.021\pm0.002$ &  $2.25\pm0.03$ &  $1094.1$ &  $10^{-232}$ \\
     $r_{\rm sat}$ &  $0.2500$ &            8 &  $0.042\pm0.003$ &  $2.25\pm0.02$ &   $583.0$ &  $10^{-122}$ \\
     $r_{\rm sat}$ &  $0.5000$ &            8 &  $0.078\pm0.009$ &  $2.22\pm0.03$ &  $1174.5$ &  $10^{-250}$ \\
     $r_{\rm sat}$ &  $1.0000$ &            8 &  $0.174\pm0.019$ &  $2.25\pm0.03$ &  $1045.2$ &  $10^{-222}$ \\
    \bottomrule
    \end{tabular}
    \label{tab:fss-F}
\end{table}

\begin{figure}
    \centering
    \includegraphics[width=\columnwidth]{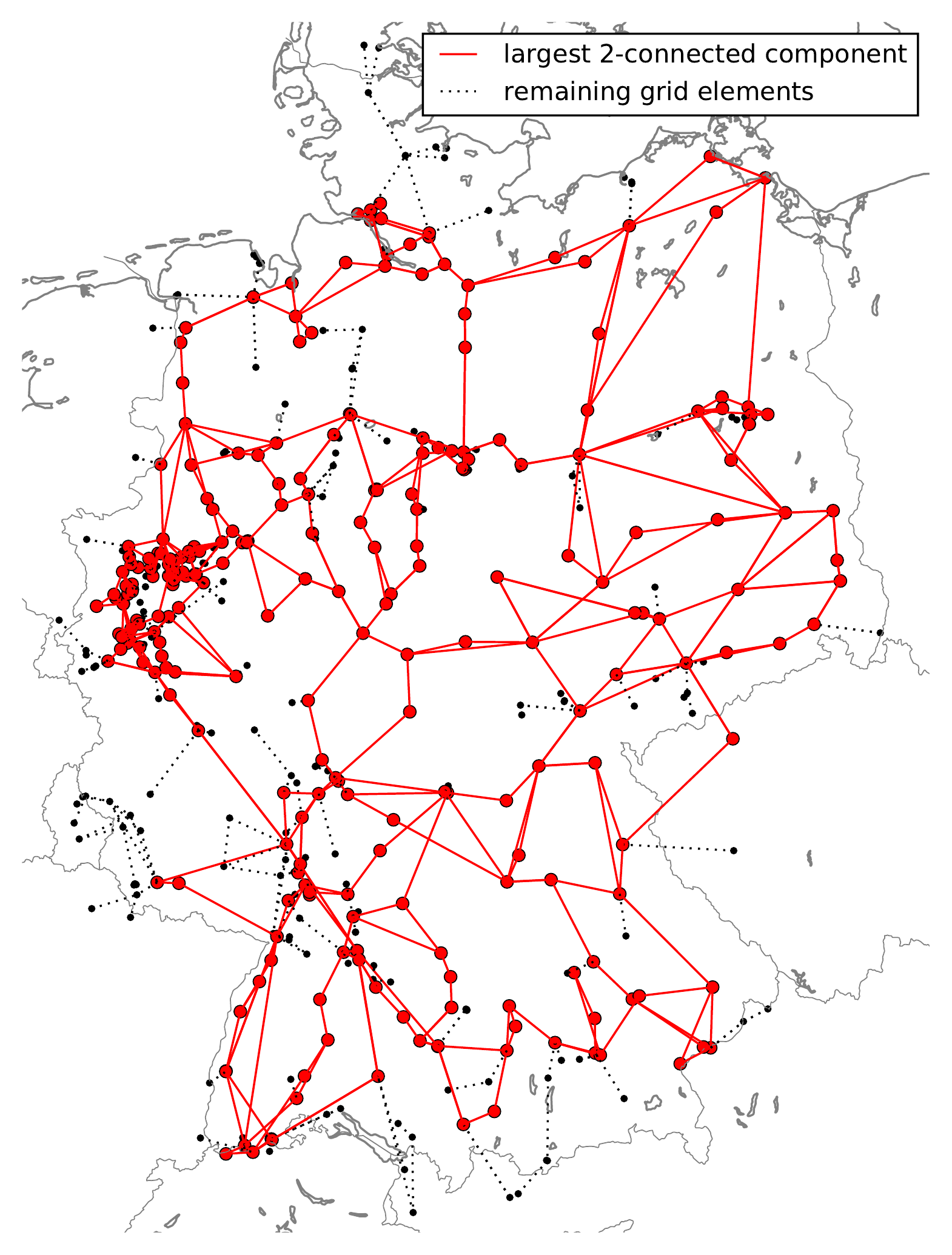}
    \vspacebeforecaption
    \caption{
        (Color online) Model for the German transmission grid (220 kV and 380
        kV), based on SciGRID data \cite{scigrid}. The largest 2-connected
        component is marked in red.
    }
    \vspaceaftercaption
    \label{fig:ger466-2c}
\end{figure}

In order to move towards more realistic grid topologies, we also study the
effect of single line additions in a model for the German transmission grid
(220 kV and 380 kV).  The model is based on open data provided by the SciGRID
project \cite{scigrid} and forms a multigraph, which means that some node pairs
are connected by more than one transmission line.  Filtering the largest
connected component, the model grid contains 467 nodes and 755 edges (see
Fig.~\ref{fig:ger466-2c}).  The graph as such is 1-connected, which means it
suffices to remove one node to render the graph disconnected.  The grid
contains stubs and also larger subgraphs that are connected by only one node to
the rest of the grid (cut nodes).  It is immediately clear that a perturbation
that changes the flow of power through the network cannot spread across such
cut nodes into another 2-connected subgraph, as there exists no alternative
path for the power to be rerouted along.  For the sake of this study, it thus
makes sense to filter only the largest 2-connected component of the grid, which
contains only 260 nodes and 479 edges (see Fig.~\ref{fig:ger466-2c}).

Despite topology, we make the same assumptions as before: We use a constant
power capacity $K_{ij} = K$ (for the lines of both voltage levels) and consider
a binary load distribution $P_i \in \{ -P, P \}$ at the nodes.  The distance
$r$ is measured by counting the edges on the shortest path to the added
transmission line which should be a reasonable approach as long as a constant
line capacity is used.

\begin{figure}
    \includegraphics[width=\columnwidth]{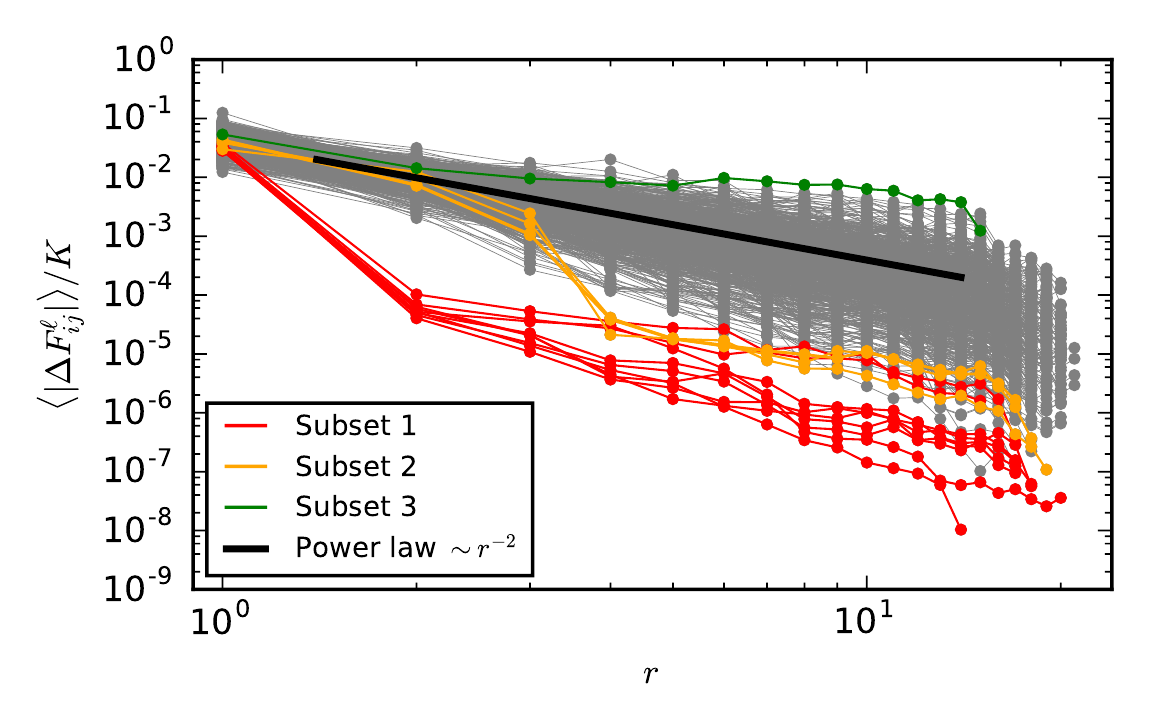}
    \vspacebeforecaption
    \caption{
        (Color online) Double-logarithmic plot of $\langle | \Delta
        F_{ij}^{\ell} | \rangle (r)$ for 880 single lines $\ell$ (gray) added
        to the largest 2-connected component of the German transmission grid
        model with $P/K = 0.25$, $w=1$ and $R=100$ \cite{scigrid}.
        For comparison, the thick black line illustrates a power law $\sim
        r^{-2}$.
        Some curves are highlighted with color, belonging to the different
        subsets of added edges defined in the text.
        The error bars correspond to a $95\,\text{\%}$ confidence level
        \cite{confidence}.
    }
    \vspaceaftercaption
    \label{fig:scaling-distance-ger}
\end{figure}

The most important difference to the cyclic 2D grid is that the resulting
response behavior will depend on the location of the added line.  As before in
the 2D square grid, we only consider adding single lines that connect
next-nearest neighbors (nodes that are two edges apart). Furthermore, we limit
their length to 208 km, which is the length of the longest line that exists in
the unperturbed grid.  There exist 880 candidates for adding a line fulfilling
these conditions. For each candidate $\ell$, we again calculate $\langle
|\Delta F_{ij}^{\ell}| \rangle(r)$, shown in
Fig.~\ref{fig:scaling-distance-ger}.

The eccentricity of each added edge can directly be read off from
Fig.~\ref{fig:scaling-distance-ger} as the maximal value of $r$ of each line.
The highest eccentricity observed is 21, which can only be realized for some
edges that are added close to the boundaries of the graph (the diameter of the
unperturbed graph is also 21).  For the majority of added lines, we find a
long-range response for $\langle|\Delta F_{ij}^{\ell}|\rangle(r)$ similar to
the 2D grid, roughly conglomerating around a power law decay with power
exponent of about 2 (compare the black line in
Fig.~\ref{fig:scaling-distance-ger}).  However, none of the curves follows a
pure power law.  Depending on the specific line added to the grid, the exponent
can be appreciably smaller or larger than 2, and the decay rate also depends on
the distance $r$.  The magnitude of $\langle|\Delta F_{ij}^{\ell}|\rangle$
varies considerably among the added lines, and sometimes shows sudden drops by
orders of magnitude at a certain distance $r$, corresponding to a large decay
rate.  In the following, we try to identify some noticable subsets of the
curves, trying to relate their characteristic long-range response behavior to
the local grid structure in the vicinity of each corresponding added line.

\begin{figure}
    \includegraphics[width=.33\columnwidth]{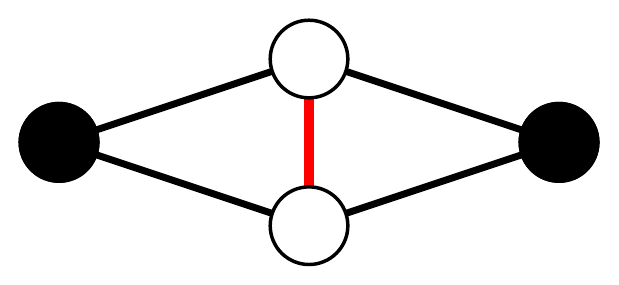}
    \vspacebeforecaption
    \caption{
        (Color online) Topological view around one of the added edges (marked
        in red) that lead to a sudden drop of absolute change of power flow in
        transmission lines of distance $r=2$ (red curves in Fig.
        \ref{fig:scaling-distance-ger}). Nodes filled with black color mark
        connecting nodes to the rest of the network.
    }
    \vspaceaftercaption
    \label{fig:edgetype1}
\end{figure}

The first considered subset of added lines corresponds to the nine red curves
in Fig.~\ref{fig:scaling-distance-ger} that drop quickly by two orders of
magnitude already at a distance of $r=2$.  It turns out that all the lines
belonging to this subset possess the same local grid structure in its vicinity,
which is illustrated by the sketch in Fig.~\ref{fig:edgetype1}.  The added line
(marked in red) perpendicularly connects two parallel routes of transmission
lines that have the same end points A and B that are the only nodes connected
to the rest of the network.  It is clear that in this case, the power flow in
the rest of the network will be altered only marginally by the addition of the
line, as it cannot contribute much to the transmission capacity of the double
route from A to B.  Apparently, it can only lead to an appreciable change of
load in the lines directly adjacent to it.

\begin{figure}
    \centering
    \includegraphics[width=\columnwidth]{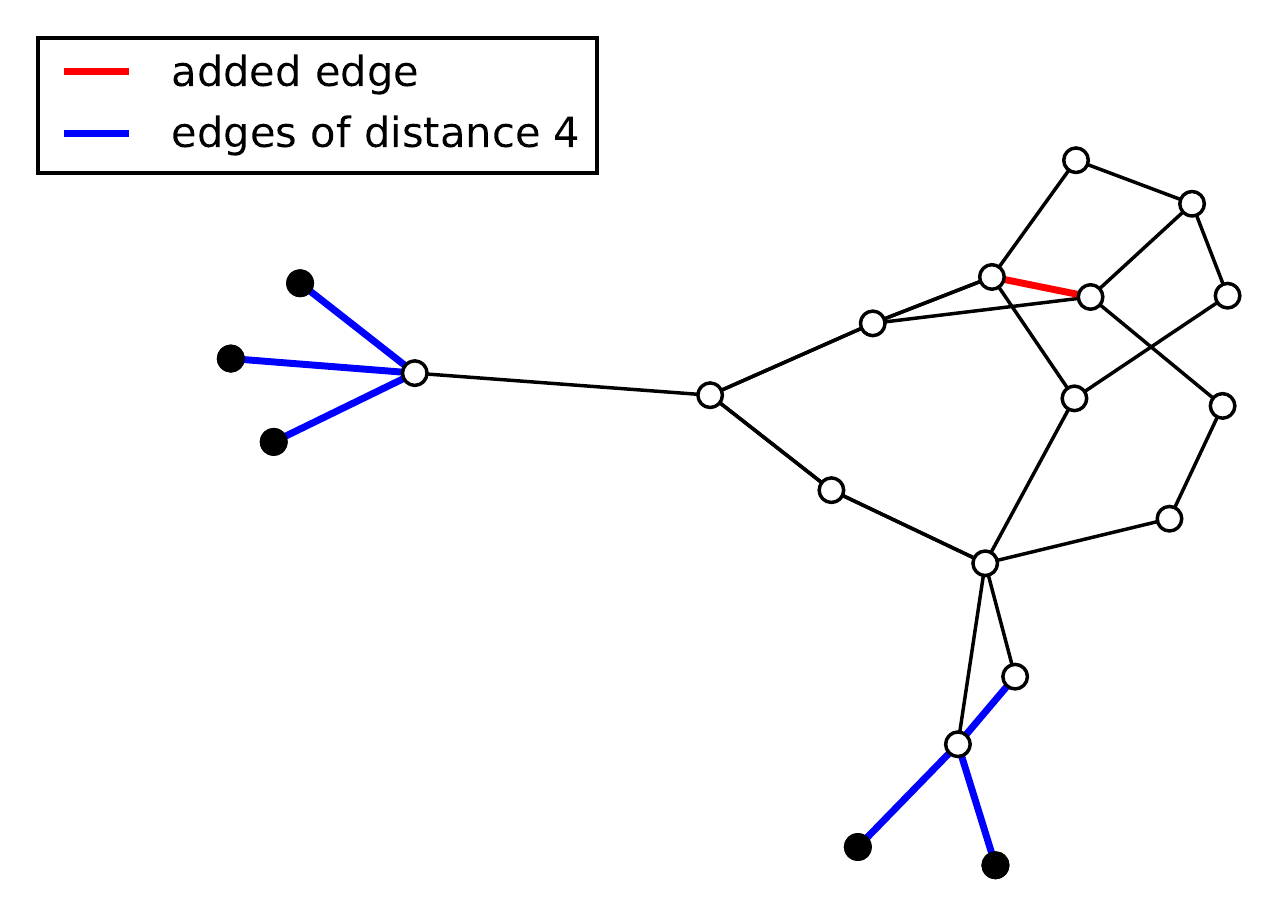}
    \caption{
        (Color online) Topological view around one of the added edges (marked
        in red) that lead to a sudden drop of absolute change of power flow in
        transmission lines of distance $r=4$ (orange curves in Fig.
        \ref{fig:scaling-distance-ger}).  Transmission lines of distance $r=4$
        are marked in blue here. Nodes filled with black color mark connecting
        nodes to the rest of the network.
    }
    \label{fig:edgetype2}
\end{figure}

The second subset we consider contains the four curves that show a sudden drop
at a distance of $r=4$, by over one order of magnitude (orange lines in
Fig.~\ref{fig:scaling-distance-ger}).  Here, we cannot identify a common local
grid structure anymore like in the example above.  However, the local grid
structure around the added lines still share some similarities: In all three
cases, the line is added inside a subgraph with a radius of about 3, that is
only weakly connected to the rest of the grid by only a small number of nodes
(two or more).  Edges of distance 4 lie just beyond these connecting nodes.  As
an example, we visualize the topology around one of the edges of this subset in
Fig.~\ref{fig:edgetype2}, with only two connecting points to the rest of the
grid.  It appears that the added transmission line considerably alters the flow
of power within the small subgraph that contains it, but beyond the few
connecting nodes, the change of power flow in the rest of the grid is moderate.

\begin{figure}
    \centering
    \includegraphics[width=\columnwidth]{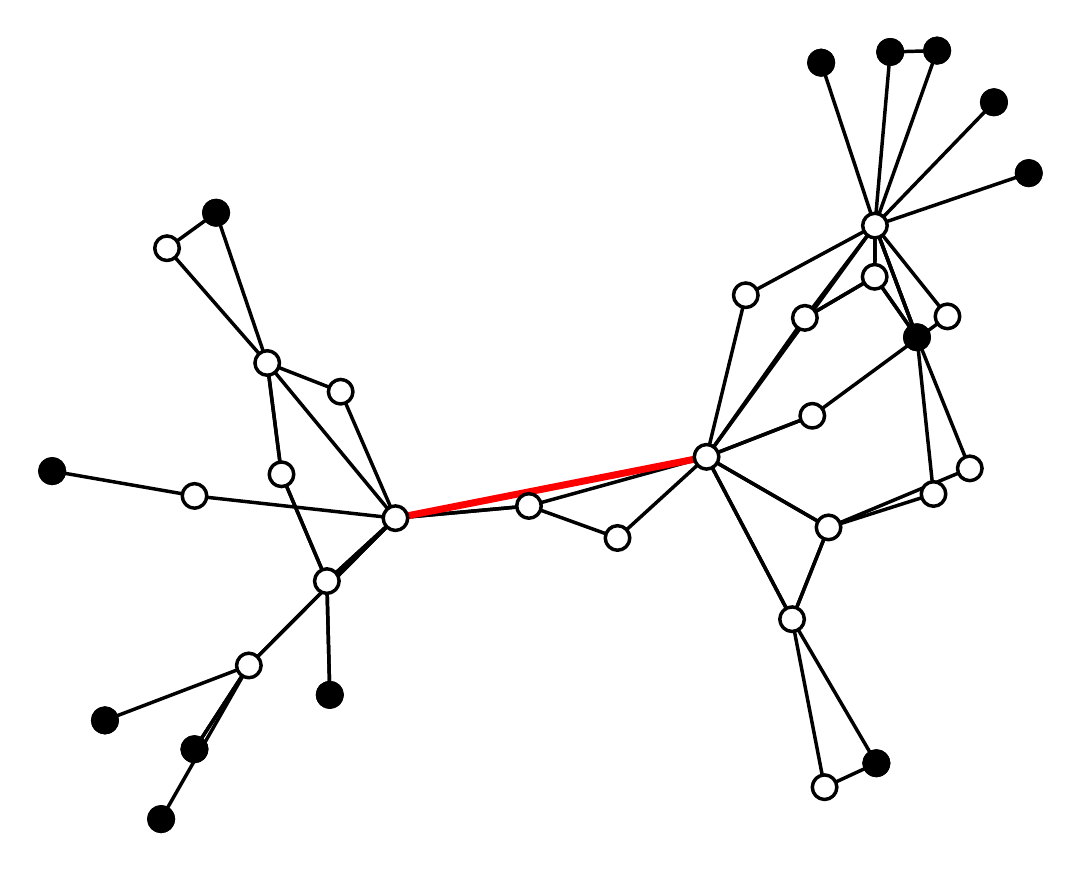}
    \caption{
        (Color online) Topological view around an added edge (marked in red)
        that leads to a profoundly weak decay of absolute power flow change
        with the distance to the added line (green curve in Fig.
        \ref{fig:scaling-distance-ger}). Nodes filled with black color mark
        connecting nodes to the rest of the network.
    }
    \label{fig:edgetype4}
\end{figure}

The curves in Fig.~\ref{fig:scaling-distance-ger} do not always decay more
strongly than a power law with exponent 2, or show drops. Some of them also
decay even slower than $\sim r^{-2}$. To gain some insight into this phenomenon
that appears to counteract the dropping mechanism, induced by changed
connectivity at certain distances, we take a look at one of the lines (marked
in green in Fig.~ \ref{fig:scaling-distance-ger}) that shows a particularly
slow decay behavior, i.e., the change in power flow hardly decreases with
distance over a wide range of distances. We illustrate the topology around this
transmission line in Fig.~\ref{fig:edgetype4}.  This particular transmission
line is quite long, close to the chosen maximum length of 208 km, and connects
two regions of high connectivity (near the cities of Cologne and Mannheim). It
runs in parallel to an existing transmission corridor and reinforces this
important connection. Qualitatively, it does not come as a surprise that the
addition of this line can influence the flow of power throughout the German
grid. Hence a small decay rate is observed in
Fig.~\ref{fig:scaling-distance-ger} for a large range of distances $r$.

All the curves (gray lines in Fig.~\ref{fig:scaling-distance-ger}), not only of
these somewhat arbitrarily chosen subsets, show drops of various magnitude at a
variety of distances to the added line.  We conclude that all of them are
triggered by the extent and nature of the connectivity of the graph at that
particular distance.  Note that in this study, the results are averaged over
all lines of a particular distance, regardless of direction.  As the structural
properties of the grid can be quite different in different regions of the grid,
a clear tendency (for example strong drops) can only be observed for some
examples in which the properties change at about the same distance in all
directions.  It can however be expected that the same mechanisms lead to the
deviations from the power law in all the curves, at all distances.  A future
study should look at this more closely, following the response behavior through
an irregular graph for each path in each direction separately.

We became aware of a recent work that examined the influence of the addition of
a single on the linear stability of an existing synchronous state
\cite{Coletta2016}.  For a simple ring topology, a classification scheme for
the increase or decrease of linear stability has been found.  It is also
demonstrated that similar effects can be observed in real power grid
topologies, considering a model for the British power grid.  It would be
interesting to further investigate the relation between the two studies.



To conclude, we have shown numerically that local grid modifications -- here in
form of the addition of a single transmission line -- cause a long-range
response in AC electricity grids.  We have identified a power law decay over a
wide range of distances to the perturbation, and quantitavely analyzed the
value of the power exponent for the case of a simple square grid topology. This
finding can also be relevant for other types of perturbations, for example
fluctuations in the power production of small, decentrally placed generators
(photovoltaics, wind power), as small changes in the generated power or small
phase perturbations \cite{Kettemann2015} can have an impact on the stable
operation of a large part of the grid.

The power exponent is found to increase with system size and saturate at a
value of $1.971 \pm 0.005$ in the medium distance regime. Further, the power
exponent decreases when approaching the critical power transmission capacity of
the system beyond which no stable state can be found anymore, and that is
distinct for a given topology and load distribution. It would be interesting to
investigate more closely how the power exponent approaches the maximal power
transmission capacity in a future study.

In real grids, like the considered model for the German transmission grid, the
response decays roughly with a power law exponent of 2 as well, but varies
strongly depending on the position of the added line within the grid and the
grid structure at each particular distance.  We find indications that the decay
rate depends on the connectivity properties of the grid in a particular
distance to the topological perturbation.  This finding underlines the
importance of this and following works on this problem, as certain local grid
structures seem to increase the decay rate of a perturbation, while others lead
to a strong decrease.

We conclude that grid regions that are only weakly connected with each other
(islanding) can effectively prevent the spreading of perturbations into other
regions of the grid, but a weak connectivity between them comes with the
expense of redundancy and hence reliability.  In the extreme case, where two
regions are connected by only one common node, no power flow changes can move
from one region to the other due to the lack of alternative routes, but its
failure is more dramatic, as the stability of the grid then depends solely on
the ability of each ``island'' to operate independently, and the n-1 criterion
is not fulfilled.

This study could be extended to other topologies like random graphs with
controllable structural properties \cite{Schultz2014}, but also to real world
topologies including realistic line parameters and load distributions.
Furthermore, the analysis could be extended to the case where the grid is not
purely inductive (with varying nodal voltages), using the complete AC power
flow equations with the reactive power $Q_i$. Ultimately, this approach could
lead to quantitative predictions if a particular grid extension measure
improves or diminishes the overall stability and controllability of power
transmission grids.

\begin{acknowledgments}

The numerical calculations have been performed using computational resources of
the Computational Laboratories for Analysis, Modeling and Visualization
(CLAMV), Jacobs University Bremen, Germany. We gratefully acknowledge support
of BMBF, CoNDyNet, FK.\ 03SF0472A.

\end{acknowledgments}




\begin{thebibliography}{99}
\bibitem{kc} For $K=K_{\rm c}$, the periodic arrangement of generators and
        consumers remains as the only possible realization, in which every
        generator is surrounded by four consumers and vise versa.
\bibitem{confidence} The error bars correspond to a confidence level of
        $95\,\text{\%}$ if they are calculated as $(\bar{Y} - 1.96 s_{\bar{Y}},
        \bar{Y} + 1.96 s_{\bar{Y}})$, where $\bar{Y}$ is the mean of the data
        and $s_{\bar{Y}}$ the standard error of the mean.
\bibitem{underestimated} The data error in our fits is solely given by the
        standard error of the mean, so an underestimation of the uncertainty in
        the raw data cannot be the cause of the bad GOF probabilities $Q$
        \cite{gof}.
\end{thebibliography}


\appendix

\begin{table*}
    \centering
    \caption{
        Fit results for regime R1.  For each parameter combination $(P,L)$,
        only data for $1 \leq r \leq 3$ is considered.  To assess the quality
        of the fit, $\chi^2$ and the quality of fit probability $Q$ are given,
        along with the number of data points $\ND$ and the number of degrees of
        freedom $\NF$ \cite{gof}.
    }
    \begin{tabular}{ccccrrrl}
    \toprule
        $P/K$ &  $L$ &                $a_1$ &            $b_1$ &  $N_{\rm D}$ &  $N_{\rm F}$ &  $\chi^2$ &                  $Q$ \\
    \midrule
     $0.0625$ &   10 &  $0.00313\pm0.00006$ &  $2.115\pm0.022$ &            3 &            1 &    $8.29$ &  $4.0\cdot 10^{ -3}$ \\
              &   20 &  $0.00294\pm0.00005$ &  $2.164\pm0.019$ &            3 &            1 &    $5.34$ &  $2.1\cdot 10^{ -2}$ \\
              &   30 &  $0.00260\pm0.00006$ &  $2.173\pm0.026$ &            3 &            1 &   $10.42$ &  $1.2\cdot 10^{ -3}$ \\
              &   40 &  $0.00231\pm0.00005$ &  $2.176\pm0.029$ &            3 &            1 &   $12.19$ &  $4.8\cdot 10^{ -4}$ \\
              &   50 &  $0.00222\pm0.00005$ &  $2.178\pm0.030$ &            3 &            1 &   $13.29$ &  $2.7\cdot 10^{ -4}$ \\
              &   60 &  $0.00199\pm0.00005$ &  $2.179\pm0.031$ &            3 &            1 &   $12.50$ &  $4.1\cdot 10^{ -4}$ \\
              &   70 &  $0.00188\pm0.00005$ &  $2.179\pm0.031$ &            3 &            1 &   $12.62$ &  $3.8\cdot 10^{ -4}$ \\
              &   80 &  $0.00196\pm0.00005$ &  $2.180\pm0.032$ &            3 &            1 &   $12.35$ &  $4.4\cdot 10^{ -4}$ \\
     $0.1250$ &   10 &  $0.00670\pm0.00012$ &  $2.115\pm0.022$ &            3 &            1 &    $8.88$ &  $2.9\cdot 10^{ -3}$ \\
              &   20 &  $0.00532\pm0.00008$ &  $2.164\pm0.019$ &            3 &            1 &    $5.06$ &  $2.4\cdot 10^{ -2}$ \\
              &   30 &  $0.00520\pm0.00011$ &  $2.173\pm0.026$ &            3 &            1 &    $9.92$ &  $1.6\cdot 10^{ -3}$ \\
              &   40 &  $0.00474\pm0.00011$ &  $2.176\pm0.029$ &            3 &            1 &   $11.59$ &  $6.6\cdot 10^{ -4}$ \\
              &   50 &  $0.00434\pm0.00011$ &  $2.178\pm0.030$ &            3 &            1 &   $12.89$ &  $3.3\cdot 10^{ -4}$ \\
              &   60 &  $0.00410\pm0.00010$ &  $2.179\pm0.031$ &            3 &            1 &   $12.73$ &  $3.6\cdot 10^{ -4}$ \\
              &   70 &  $0.00371\pm0.00009$ &  $2.179\pm0.031$ &            3 &            1 &   $13.25$ &  $2.7\cdot 10^{ -4}$ \\
              &   80 &  $0.00379\pm0.00010$ &  $2.179\pm0.032$ &            3 &            1 &   $13.74$ &  $2.1\cdot 10^{ -4}$ \\
     $0.2500$ &   10 &  $0.01269\pm0.00023$ &  $2.114\pm0.022$ &            3 &            1 &    $7.89$ &  $5.0\cdot 10^{ -3}$ \\
              &   20 &  $0.01129\pm0.00017$ &  $2.164\pm0.019$ &            3 &            1 &    $5.35$ &  $2.1\cdot 10^{ -2}$ \\
              &   30 &  $0.01015\pm0.00021$ &  $2.173\pm0.026$ &            3 &            1 &    $9.81$ &  $1.7\cdot 10^{ -3}$ \\
              &   40 &  $0.00899\pm0.00021$ &  $2.176\pm0.029$ &            3 &            1 &   $11.46$ &  $7.1\cdot 10^{ -4}$ \\
              &   50 &  $0.00864\pm0.00021$ &  $2.178\pm0.030$ &            3 &            1 &   $12.70$ &  $3.7\cdot 10^{ -4}$ \\
              &   60 &  $0.00791\pm0.00019$ &  $2.179\pm0.031$ &            3 &            1 &   $13.09$ &  $3.0\cdot 10^{ -4}$ \\
              &   70 &  $0.00776\pm0.00019$ &  $2.179\pm0.031$ &            3 &            1 &   $13.54$ &  $2.3\cdot 10^{ -4}$ \\
              &   80 &  $0.00727\pm0.00018$ &  $2.180\pm0.031$ &            3 &            1 &   $11.78$ &  $6.0\cdot 10^{ -4}$ \\
     $0.5000$ &   10 &  $0.02501\pm0.00047$ &  $2.114\pm0.023$ &            3 &            1 &    $8.42$ &  $3.7\cdot 10^{ -3}$ \\
              &   20 &  $0.02283\pm0.00033$ &  $2.163\pm0.018$ &            3 &            1 &    $4.47$ &  $3.4\cdot 10^{ -2}$ \\
              &   30 &  $0.02022\pm0.00041$ &  $2.172\pm0.025$ &            3 &            1 &    $8.67$ &  $3.2\cdot 10^{ -3}$ \\
              &   40 &  $0.01839\pm0.00041$ &  $2.176\pm0.028$ &            3 &            1 &   $11.01$ &  $9.0\cdot 10^{ -4}$ \\
              &   50 &  $0.01851\pm0.00043$ &  $2.177\pm0.029$ &            3 &            1 &   $11.61$ &  $6.6\cdot 10^{ -4}$ \\
              &   60 &  $0.01611\pm0.00039$ &  $2.179\pm0.030$ &            3 &            1 &   $11.86$ &  $5.7\cdot 10^{ -4}$ \\
              &   70 &  $0.01545\pm0.00038$ &  $2.179\pm0.030$ &            3 &            1 &   $11.95$ &  $5.5\cdot 10^{ -4}$ \\
              &   80 &  $0.01553\pm0.00038$ &  $2.179\pm0.031$ &            3 &            1 &   $12.66$ &  $3.7\cdot 10^{ -4}$ \\
     $1.0000$ &   10 &  $0.05111\pm0.00116$ &  $2.108\pm0.028$ &            3 &            1 &   $13.01$ &  $3.1\cdot 10^{ -4}$ \\
              &   20 &  $0.04550\pm0.00027$ &  $2.152\pm0.007$ &            3 &            1 &    $0.83$ &  $3.6\cdot 10^{ -1}$ \\
              &   30 &  $0.04073\pm0.00052$ &  $2.164\pm0.016$ &            3 &            1 &    $3.85$ &  $5.0\cdot 10^{ -2}$ \\
              &   40 &  $0.03653\pm0.00056$ &  $2.170\pm0.019$ &            3 &            1 &    $5.27$ &  $2.2\cdot 10^{ -2}$ \\
              &   50 &  $0.03501\pm0.00053$ &  $2.169\pm0.019$ &            3 &            1 &    $4.69$ &  $3.0\cdot 10^{ -2}$ \\
              &   60 &  $0.03369\pm0.00059$ &  $2.171\pm0.022$ &            3 &            1 &    $5.87$ &  $1.5\cdot 10^{ -2}$ \\
              &   70 &  $0.03185\pm0.00061$ &  $2.175\pm0.024$ &            3 &            1 &    $7.81$ &  $5.2\cdot 10^{ -3}$ \\
              &   80 &  $0.02897\pm0.00056$ &  $2.176\pm0.024$ &            3 &            1 &    $7.80$ &  $5.2\cdot 10^{ -3}$ \\
    \bottomrule
    \end{tabular}
    \label{tab:scaling1}
\end{table*}

\begin{table*}
    \centering
    \caption{
        Fit results for regime R2. For each parameter combination $(P,L)$, only
        data for $4 \leq r \leq (2L/5)-1$ is considered. To assess the quality
        of the fit, $\chi^2$ and the quality of fit probability $Q$ are given,
        along with the number of data points $\ND$ and the number of degrees of
        freedom $\NF$ \cite{gof}.
    }
    \begin{tabular}{ccccrrrl}
    \toprule
        $P/K$ &  $L$ &                $a_2$ &            $b_2$ &  $N_{\rm D}$ &  $N_{\rm F}$ &  $\chi^2$ &                   $Q$ \\
    \midrule
     $0.0625$ &   20 &  $0.00252\pm0.00008$ &  $1.960\pm0.017$ &            4 &            2 &    $7.56$ &   $2.3\cdot 10^{ -2}$ \\
              &   30 &  $0.00222\pm0.00006$ &  $1.966\pm0.012$ &            8 &            6 &   $67.04$ &   $1.7\cdot 10^{-12}$ \\
              &   40 &  $0.00195\pm0.00004$ &  $1.967\pm0.009$ &           12 &           10 &  $157.89$ &   $8.8\cdot 10^{-29}$ \\
              &   50 &  $0.00187\pm0.00003$ &  $1.969\pm0.007$ &           16 &           14 &  $257.88$ &   $6.7\cdot 10^{-47}$ \\
              &   60 &  $0.00166\pm0.00003$ &  $1.969\pm0.006$ &           20 &           18 &  $368.69$ &   $3.0\cdot 10^{-67}$ \\
              &   70 &  $0.00157\pm0.00002$ &  $1.970\pm0.005$ &           24 &           22 &  $509.90$ &   $6.3\cdot 10^{-94}$ \\
              &   80 &  $0.00163\pm0.00002$ &  $1.971\pm0.005$ &           28 &           26 &  $669.94$ &  $1.4\cdot 10^{-124}$ \\
     $0.1250$ &   20 &  $0.00457\pm0.00014$ &  $1.960\pm0.017$ &            4 &            2 &    $7.21$ &   $2.7\cdot 10^{ -2}$ \\
              &   30 &  $0.00443\pm0.00011$ &  $1.966\pm0.012$ &            8 &            6 &   $64.22$ &   $6.2\cdot 10^{-12}$ \\
              &   40 &  $0.00400\pm0.00009$ &  $1.967\pm0.009$ &           12 &           10 &  $151.00$ &   $2.3\cdot 10^{-27}$ \\
              &   50 &  $0.00365\pm0.00007$ &  $1.969\pm0.007$ &           16 &           14 &  $250.96$ &   $1.8\cdot 10^{-45}$ \\
              &   60 &  $0.00344\pm0.00006$ &  $1.969\pm0.006$ &           20 &           18 &  $376.49$ &   $7.2\cdot 10^{-69}$ \\
              &   70 &  $0.00310\pm0.00005$ &  $1.970\pm0.005$ &           24 &           22 &  $534.89$ &   $3.8\cdot 10^{-99}$ \\
              &   80 &  $0.00316\pm0.00005$ &  $1.971\pm0.005$ &           28 &           26 &  $743.00$ &  $6.8\cdot 10^{-140}$ \\
     $0.2500$ &   20 &  $0.00969\pm0.00029$ &  $1.960\pm0.017$ &            4 &            2 &    $7.68$ &   $2.2\cdot 10^{ -2}$ \\
              &   30 &  $0.00864\pm0.00022$ &  $1.966\pm0.012$ &            8 &            6 &   $63.89$ &   $7.3\cdot 10^{-12}$ \\
              &   40 &  $0.00759\pm0.00016$ &  $1.967\pm0.009$ &           12 &           10 &  $150.37$ &   $3.1\cdot 10^{-27}$ \\
              &   50 &  $0.00727\pm0.00013$ &  $1.969\pm0.007$ &           16 &           14 &  $250.56$ &   $2.2\cdot 10^{-45}$ \\
              &   60 &  $0.00663\pm0.00011$ &  $1.969\pm0.006$ &           20 &           18 &  $389.03$ &   $1.8\cdot 10^{-71}$ \\
              &   70 &  $0.00649\pm0.00010$ &  $1.970\pm0.005$ &           24 &           22 &  $549.15$ &  $3.9\cdot 10^{-102}$ \\
              &   80 &  $0.00606\pm0.00009$ &  $1.971\pm0.005$ &           28 &           26 &  $644.71$ &  $2.7\cdot 10^{-119}$ \\
     $0.5000$ &   20 &  $0.01961\pm0.00058$ &  $1.959\pm0.017$ &            4 &            2 &    $6.98$ &   $3.0\cdot 10^{ -2}$ \\
              &   30 &  $0.01722\pm0.00043$ &  $1.965\pm0.012$ &            8 &            6 &   $58.99$ &   $7.2\cdot 10^{-11}$ \\
              &   40 &  $0.01552\pm0.00033$ &  $1.966\pm0.009$ &           12 &           10 &  $152.97$ &   $9.1\cdot 10^{-28}$ \\
              &   50 &  $0.01559\pm0.00028$ &  $1.968\pm0.007$ &           16 &           14 &  $240.61$ &   $2.5\cdot 10^{-43}$ \\
              &   60 &  $0.01351\pm0.00023$ &  $1.969\pm0.006$ &           20 &           18 &  $365.40$ &   $1.4\cdot 10^{-66}$ \\
              &   70 &  $0.01292\pm0.00020$ &  $1.970\pm0.005$ &           24 &           22 &  $507.47$ &   $2.0\cdot 10^{-93}$ \\
              &   80 &  $0.01296\pm0.00019$ &  $1.971\pm0.005$ &           28 &           26 &  $717.61$ &  $1.5\cdot 10^{-134}$ \\
     $1.0000$ &   20 &  $0.03892\pm0.00115$ &  $1.953\pm0.017$ &            4 &            2 &    $7.09$ &   $2.9\cdot 10^{ -2}$ \\
              &   30 &  $0.03467\pm0.00081$ &  $1.961\pm0.011$ &            8 &            6 &   $56.20$ &   $2.7\cdot 10^{-10}$ \\
              &   40 &  $0.03066\pm0.00064$ &  $1.961\pm0.009$ &           12 &           10 &  $145.92$ &   $2.6\cdot 10^{-26}$ \\
              &   50 &  $0.02966\pm0.00052$ &  $1.967\pm0.007$ &           16 &           14 &  $212.28$ &   $1.7\cdot 10^{-37}$ \\
              &   60 &  $0.02837\pm0.00046$ &  $1.968\pm0.006$ &           20 &           18 &  $321.22$ &   $2.0\cdot 10^{-57}$ \\
              &   70 &  $0.02672\pm0.00041$ &  $1.969\pm0.005$ &           24 &           22 &  $508.00$ &   $1.6\cdot 10^{-93}$ \\
              &   80 &  $0.02422\pm0.00035$ &  $1.970\pm0.005$ &           28 &           26 &  $675.52$ &  $9.8\cdot 10^{-126}$ \\
    \bottomrule
    \end{tabular}
    \label{tab:scaling2}
\end{table*}

\begin{table*}
    \centering
    \caption{
        Fit results for regime R3. For each parameter combination $(P,L)$, only
        data for $(2L/5)+1 \leq r \leq 7L/10$ is considered. To assess the
        quality of the fit, $\chi^2$ and the quality of fit probability $Q$ are
        given, along with the number of data points $\ND$ and the number of
        degrees of freedom $\NF$ \cite{gof}.
    }
    \begin{tabular}{lrllrrrl}
    \toprule
        $P/K$ &  $L$ &                      $g$ &                $h$ &  $\ND$ &  $\NF$ &  $\chi^2$ &                   $Q$ \\
    \midrule
     $0.0625$ &   20 &  $0.0001712\pm0.0000050$ &  $0.1721\pm0.0026$ &      6 &      4 &   $44.64$ &   $4.7\cdot 10^{ -9}$ \\
              &   30 &  $0.0000652\pm0.0000016$ &  $0.1126\pm0.0015$ &      9 &      7 &  $127.30$ &   $2.3\cdot 10^{-24}$ \\
              &   40 &  $0.0000323\pm0.0000007$ &  $0.0841\pm0.0010$ &     12 &     10 &  $233.59$ &   $1.5\cdot 10^{-44}$ \\
              &   50 &  $0.0000197\pm0.0000003$ &  $0.0670\pm0.0006$ &     15 &     13 &  $344.11$ &   $1.3\cdot 10^{-65}$ \\
              &   60 &  $0.0000122\pm0.0000002$ &  $0.0557\pm0.0005$ &     18 &     16 &  $428.93$ &   $3.1\cdot 10^{-81}$ \\
              &   70 &  $0.0000085\pm0.0000001$ &  $0.0476\pm0.0004$ &     21 &     19 &  $562.26$ &  $4.6\cdot 10^{-107}$ \\
              &   80 &  $0.0000067\pm0.0000001$ &  $0.0416\pm0.0003$ &     24 &     22 &  $701.53$ &  $3.7\cdot 10^{-134}$ \\
     $0.1250$ &   20 &  $0.0003102\pm0.0000091$ &  $0.1721\pm0.0026$ &      6 &      4 &   $42.70$ &   $1.2\cdot 10^{ -8}$ \\
              &   30 &  $0.0001303\pm0.0000032$ &  $0.1126\pm0.0015$ &      9 &      7 &  $122.30$ &   $2.5\cdot 10^{-23}$ \\
              &   40 &  $0.0000663\pm0.0000014$ &  $0.0841\pm0.0010$ &     12 &     10 &  $223.95$ &   $1.6\cdot 10^{-42}$ \\
              &   50 &  $0.0000386\pm0.0000007$ &  $0.0670\pm0.0006$ &     15 &     13 &  $335.59$ &   $8.3\cdot 10^{-64}$ \\
              &   60 &  $0.0000252\pm0.0000004$ &  $0.0557\pm0.0005$ &     18 &     16 &  $437.60$ &   $4.7\cdot 10^{-83}$ \\
              &   70 &  $0.0000167\pm0.0000002$ &  $0.0476\pm0.0004$ &     21 &     19 &  $588.51$ &  $1.3\cdot 10^{-112}$ \\
              &   80 &  $0.0000130\pm0.0000002$ &  $0.0416\pm0.0003$ &     24 &     22 &  $774.04$ &  $1.8\cdot 10^{-149}$ \\
     $0.2500$ &   20 &  $0.0006579\pm0.0000192$ &  $0.1721\pm0.0026$ &      6 &      4 &   $45.52$ &   $3.1\cdot 10^{ -9}$ \\
              &   30 &  $0.0002542\pm0.0000062$ &  $0.1126\pm0.0015$ &      9 &      7 &  $122.28$ &   $2.6\cdot 10^{-23}$ \\
              &   40 &  $0.0001258\pm0.0000026$ &  $0.0841\pm0.0010$ &     12 &     10 &  $223.18$ &   $2.3\cdot 10^{-42}$ \\
              &   50 &  $0.0000767\pm0.0000014$ &  $0.0670\pm0.0006$ &     15 &     13 &  $334.95$ &   $1.1\cdot 10^{-63}$ \\
              &   60 &  $0.0000485\pm0.0000008$ &  $0.0557\pm0.0005$ &     18 &     16 &  $451.21$ &   $6.4\cdot 10^{-86}$ \\
              &   70 &  $0.0000349\pm0.0000005$ &  $0.0476\pm0.0004$ &     21 &     19 &  $603.20$ &  $1.1\cdot 10^{-115}$ \\
              &   80 &  $0.0000250\pm0.0000003$ &  $0.0416\pm0.0003$ &     24 &     22 &  $677.87$ &  $3.6\cdot 10^{-129}$ \\
     $0.5000$ &   20 &  $0.0013324\pm0.0000391$ &  $0.1721\pm0.0026$ &      6 &      4 &   $43.12$ &   $9.8\cdot 10^{ -9}$ \\
              &   30 &  $0.0005070\pm0.0000125$ &  $0.1126\pm0.0015$ &      9 &      7 &  $115.94$ &   $5.4\cdot 10^{-22}$ \\
              &   40 &  $0.0002573\pm0.0000054$ &  $0.0841\pm0.0010$ &     12 &     10 &  $227.80$ &   $2.5\cdot 10^{-43}$ \\
              &   50 &  $0.0001646\pm0.0000029$ &  $0.0670\pm0.0006$ &     15 &     13 &  $321.88$ &   $6.3\cdot 10^{-61}$ \\
              &   60 &  $0.0000991\pm0.0000015$ &  $0.0557\pm0.0005$ &     18 &     16 &  $425.94$ &   $1.3\cdot 10^{-80}$ \\
              &   70 &  $0.0000695\pm0.0000010$ &  $0.0476\pm0.0004$ &     21 &     19 &  $558.25$ &  $3.2\cdot 10^{-106}$ \\
              &   80 &  $0.0000534\pm0.0000007$ &  $0.0416\pm0.0003$ &     24 &     22 &  $752.61$ &  $6.0\cdot 10^{-145}$ \\
     $1.0000$ &   20 &  $0.0026634\pm0.0000854$ &  $0.1717\pm0.0029$ &      6 &      4 &   $51.61$ &   $1.7\cdot 10^{-10}$ \\
              &   30 &  $0.0010281\pm0.0000263$ &  $0.1125\pm0.0015$ &      9 &      7 &  $135.50$ &   $4.4\cdot 10^{-26}$ \\
              &   40 &  $0.0004599\pm0.0000090$ &  $0.0779\pm0.0010$ &     12 &     10 &   $60.52$ &   $2.9\cdot 10^{ -9}$ \\
              &   50 &  $0.0003145\pm0.0000054$ &  $0.0670\pm0.0006$ &     15 &     13 &  $294.22$ &   $3.9\cdot 10^{-55}$ \\
              &   60 &  $0.0002085\pm0.0000032$ &  $0.0556\pm0.0005$ &     18 &     16 &  $389.78$ &   $5.0\cdot 10^{-73}$ \\
              &   70 &  $0.0001441\pm0.0000020$ &  $0.0476\pm0.0004$ &     21 &     19 &  $571.52$ &  $5.1\cdot 10^{-109}$ \\
              &   80 &  $0.0001001\pm0.0000013$ &  $0.0416\pm0.0003$ &     24 &     22 &  $729.50$ &  $4.6\cdot 10^{-140}$ \\
    \bottomrule
    \end{tabular}
    \label{tab:scaling3}
\end{table*}

\end{document}